\documentclass{cernyrep}

\begin{document}

\title{Magnetic measurement with coils and wires}
\author{L. Walckiers}
\institute{CERN, Geneva, Switzerland}

% \tableofcontents 

\maketitle % genere badbox

%Abstract.tex

\begin{abstract}
Accelerator magnets steer particle beams according to the field
integrated along the trajectory over the magnet length. Purpose-wound
coils measure these relevant parameters with high precision and
complement efficiently point-like measurements performed with Hall
plates or NMR probes. The rotating coil method gives a complete
two-dimensional description of the magnetic field in a series of
normal and skew multipoles. The more recent single stretched wire is a
reference instrument to measure field integrals and to find the
magnetic axis.
\end{abstract}

% Introduction.tex

\section{Introduction} \label{sec:Introduction}

The field of measurement of accelerator magnets has followed the
requirements dictated by developments in accelerator
technology. Synchrotron Light Sources require stringent magnetic axis
alignment. The field quality of superconducting magnets for an
accelerator like the Large Hadron Collider (LHC) has been measured to
unprecedented precision, including the time and ramp rate dependent
effects due to superconductor cables.

Measuring coils and more recently stretched wires are used extensively
for accelerator magnets since they measure the parameters seen by
particle beams: magnetic field components in the plane perpendicular
to their trajectory and integrated over the length of the
magnets. Coils and wires allow fast measurements with data already
reduced to the requirements.

This field of physics has benefited from the development in electronic
components. Fast acquisition voltmeters or recent 18-bit ADCs with
sampling times in the range of microseconds increase the bandwidth and
the precision of voltage integrated over time, the basis of this type
of measurement.

Section \ref{sec:dipolecoil} describes the classical method of coils
flipped by half a turn inside dipole magnets or static coils in pulsed
magnets. They are still in use to measure resistive dipoles having
flat horizontal aperture and in particular for magnets having a small
radius of curvature. In addition, the theory developed for these
simple coils must be understood to avoid flaws with more sophisticated
measurement methods.  Section \ref{sec:Coilsarray} introduces the
concept of coil arrays to suppress the contribution from some harmonic
components, a concept fundamental to the accurate use of the harmonic
coil method.

Sections \ref{sec:sswdipole} to \ref{sec:Vibrating} group the Single
Stretched Wire (SSW) based methods. They measure with high absolute
precision the fundamental parameters of accelerator magnets.  Section
\ref{sec:sswdipole} introduces the equipment and demonstrates that the
SSW gives an absolute value of the field integrated over the magnet
length with a minimum of calibration concern.  Section
\ref{sec:sswaxis} details how to reference in all dimensions the
position of the quadrupole magnet axis and field direction.  Section
\ref{sec:sswquadrupole} gives the most accurate method to quantify the
integrated field gradient and address the issues related to wire
deflection due to gravity and magnetic susceptibility.  The vibrating
wire method described in Section~\ref{sec:Vibrating} is still subject
to interesting developments in order to find the axis of individual
magnets aligned on a girder. It is probably the only candidate method
to measure small-aperture magnets under development for high-energy
linear accelerators.

Sections \ref{sec:Harmcoil} to \ref{sec:HarmExperience} cover the
harmonic or rotating coil method. This technique gives high resolution
and measures in one coil revolution all relevant parameters of any
accelerator magnet. Both theoretical and experimental developments
allow one to confidently design sophisticated instruments measuring
with high bandwidth and precision the full harmonic content of a
magnet.
% Coils.tex
\section{Coils to measure dipole magnets}
\label{sec:dipolecoil}

The following section describes methods that measure only partially
the 2D field along the axis of accelerator magnets. The rotating coil
method (Section~\ref{sec:Harmcoil} and following) gives a complete
measurement of any 2D field expressed in a formal way by a series of
complex multipoles. It cannot, however, be used in the following cases
which are relevant for most `accelerator magnets' compared to storage
rings:

\begin{itemize}
\item dipole magnets of small accelerators are bent,
\item they mostly have wide horizontal apertures compared to the gap height,
\item the rotating coil method does not (easily) measure pulsed
      fields, \ie{} when $\Delta B / \Delta t$ cannot be neglected
      over one coil revolution period.
\end{itemize}

\subsection{Flip coils for dipole magnet strength}
\label{sec:flip}

\begin{figure}
\centering
\includegraphics[width=10cm]{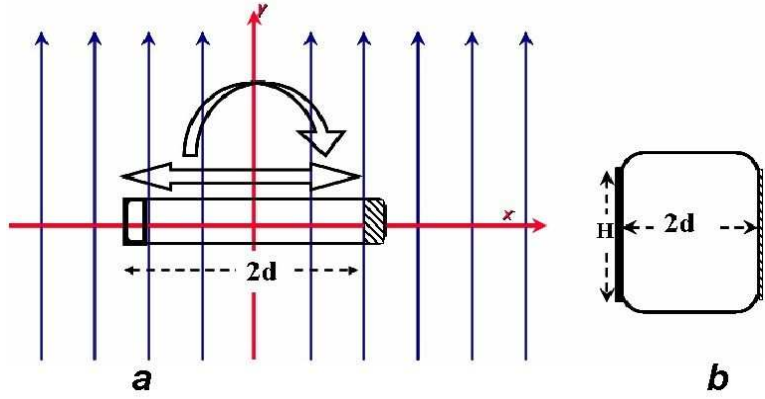} 
\caption[]{(a) A simple coil, flipped in a dipole field to measure the
           central field, displaced laterally to measure the field
           quality. (b) Square coil with a single layer winding
           insensitive to the sextupole terms when $H = 2d$.}
\label{fig:flipcoil}
\end{figure}

The flux picked by the single-turn coil of width $2d$ sketched in
\Fref{fig:flipcoil}(a) that is longer than the magnet of length $L$
and that rotates by half a turn is

\begin{equation}
\Psi(\pi)-\Psi(0)= 2 \cdot \int_{0}^{L}{\int_{-d}^{d}{B_y(x) \cdot dx \cdot dl}} ~. 
\label{eq:coil1}   
\end{equation}

By assuming either a perfect dipole, i.e., $B_y(x,y)$ constant over
the aperture or a coil of width $2d$ small compared to the field
errors, i.e., the higher harmonics present in the magnet, the
quantities relevant for the particle beam are deduced for the
excitation current $I$ in the magnet assumed to be constant during the
time to flip the coil:

\begin{align}
\text{Dipole strength} & 
     = \left[\int{B_y \cdot dl}\right](I)\qquad\text{in~~[T$\cdot$m].}
\label{eq:coil2} \\
\text{Transfer function} &
     = \left[\int{B_y \cdot dl}\right](I)/I\qquad\text{in~~[T$\cdot$m{}/{}A].}
\label{eq:coil3}   
\end{align}

The measuring coil has to be longer than the magnet. A rule of thumb
says that the coil should extend outside both magnet ends by 2.5 times
the aperture. To quantify the validity of this approximation, there is
no other way than to perform a $B(z)$ scanning with a Hall plate or to
make a full 3D calculation of the stray field in the magnet ends.

\subsection{Coils displaced horizontally to measure the field quality}
\label{sec:coils-displaced}

The first order imperfections to consider in a resistive magnet are
the variation of the vertical field over the aperture width $\Delta
B_y(x)$. The situation is more complex in superconducting magnets
where a non-negligible horizontal field component $B_x(x)$ can be
present in the horizontal mid-plane. The full 2D complex formalism has
then to be applied to describe the field harmonics that can be
measured by the rotating coil method covered from \Sref{sec:Harmcoil}
onward. An horizontal displacement $\Delta x$ of the coil of
\Fref{fig:flipcoil}(a) leads per unit length along the magnet axis to
 
 \begin{equation}
\frac{\Delta B_y(x)}{\Delta x}= \frac{\Psi(x+\Delta x)-\Psi(x)}{2d}~. 
\label{eq:coil4}   
\end{equation}

\Fref[b]{fig:cnaocurves} gives the field quality in the horizontal
symmetry plane of 10 dipole magnets for the CNAO facility
(\Sref{sec:CNAO}). These curves were measured, for efficiency reasons,
in pulsed current mode by 11 different static coils rather than 10
horizontal displacements.

\begin{figure}
\centering
\includegraphics[width=12cm]{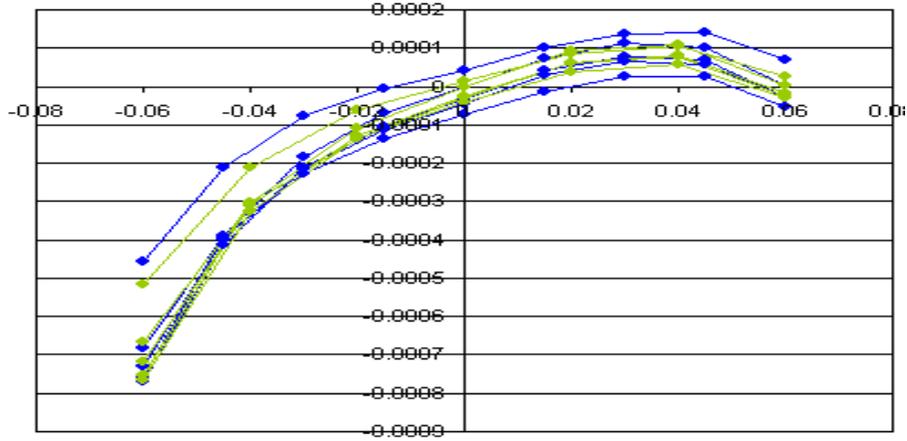} 
\caption[]{$\Delta B_y(x)/\Delta x$ measured on 10 CNAO dipole magnets
           with the 11-coil fluxmeter}
\label{fig:cnaocurves}
\end{figure}

\subsection{A real coil has finite dimensions for the winding}
\label{sec:realcoil}

One way to increase the voltage amplitude at the coil terminals is to
increase the width $2d$ therefore picking a vertical field
non-constant over this width according to the field quality of the
magnet. What is measured is no longer the field at the centre of the
coil. The real part of Eq. (\ref{eq:eqH6}) gives

\begin{equation}
B_y(x,y)= B_1 + B_2 \cdot x +B_3 \cdot (x^2-y2) + ...~. 
\label{eq:coil5}   
\end{equation}

Integrating over the coil width going from $[x-d]$ to $[x+d]$ leads,
with the hypothesis of a negligible winding height, to

\begin{equation}
\Psi(x)= B_1 \cdot 2d +B_2 \cdot  2dx +B_3 \cdot (6dx^2 + 2d^3) + ...~.
\label{eq:coil51}   
\end{equation}

This equation indicates that a measurement with a half-turn flip is
never sensitive to the even harmonics (quadrupole, octupole, etc.) as
long as the axis of rotation is centred in the middle of the coil. On
the other hand, the sextupole component enters in a dipole strength
measurement:

\begin{equation}
\Psi(\pi,x=0)- \Psi(0,x=0)= 2.(B_1 \cdot 2d  +B_3 \cdot  2d^3) + ...~.
\label{eq:coil6}   
\end{equation}

All multipole terms enter with different and varying sensitivities
when measuring the field quality by a displaced coil:

\begin{equation}
\Psi(x+\delta)-\Psi(x)= 
  B_2 \cdot 2d\delta +B_3 \cdot 6d\cdot(2x\delta + \delta^2) + ...~.
\label{eq:coil7}   
\end{equation}

Coils with several turns are used to further increase the
sensitivity. The winding has then a non-zero cross-section. The coil
of \Fref{fig:flipcoil}(b), with coil width equal to the winding height
is easily calculated to have no sensitivity to the sextupole $B_3$
terms. The issue detailed by Eq.~(\ref{eq:coil6}) for flip coil
measurement is therefore limited to higher orders: starting with the
decapole term present in the magnet.

\begin{figure}
\centering
\includegraphics[width=88mm]{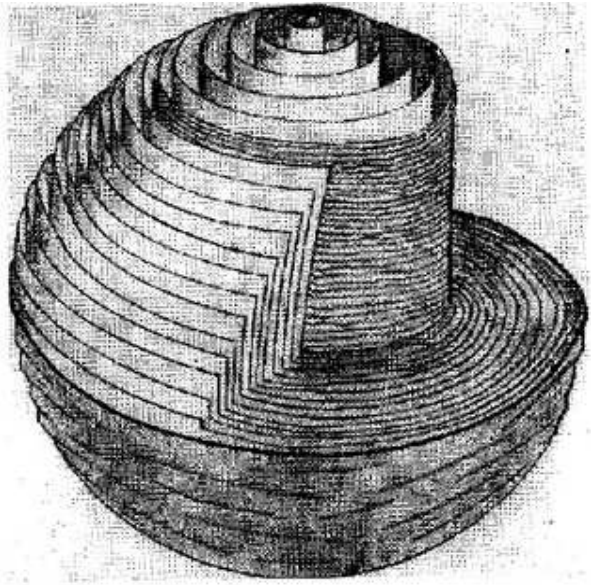} 
\caption[]{A fluxball, composed of cylindrical windings, measures the
           field at the central point of the ball}
\label{fig:fluxball}
\end{figure}

This way of avoiding sensitivity to higher multipole terms was
generalized in theory long ago with the fluxball~\cite{Brown45}. This
fluxball coil, as sketched in \Fref{fig:fluxball} seems difficult to
manufacture but picks up a flux equal to the field at the centre of
the ball and independent of the spatial harmonics present. A coil
approximating this possibility of measuring the field central to a
highly sensitive coil probe has been proposed in \Bref{GreenCAS98}.

\subsection{Static coils in pulsed fields}

Most accelerator magnets have to be measured in fast current ramping
conditions. The coil of \Fref{fig:flipcoil}(b), insensitive in
that case to terms lower than the decapole, is an easy tool to use in
fixed position in a field pulsed from 0 to nominal value. Modern
integrators with large bandwidth and time resolution connected to such
a coil can give the full $B_1(t)$ curve and measure saturation effects
of the iron yoke. Hysteresis and eddy current effects can be separated
by measurements at different ramp rates.  One precaution deals with
the remanent field of the magnet being measured, i.e., the field at
zero current value. Three~ways can be used to solve it:
\begin{itemize}
\item have a bipolar power supply to perform symmetric sweeps from
      negative to positive maximum current;
\item demagnetize the magnet first, with either a bipolar supply or a
      supply with an inverter;
\item measure the remanent field with a flip coil or Hall plates for
      instance, low accuracy is sufficient in most of the cases.
\end{itemize}
% Coilsarray.tex
\section{Arrays of coils to measure quadrupoles or higher order multipoles}
\label{sec:Coilsarray}

\subsection{Two-coil array for quadrupole strength and field quality}

\begin{figure}
\centering
\includegraphics[width=9cm]{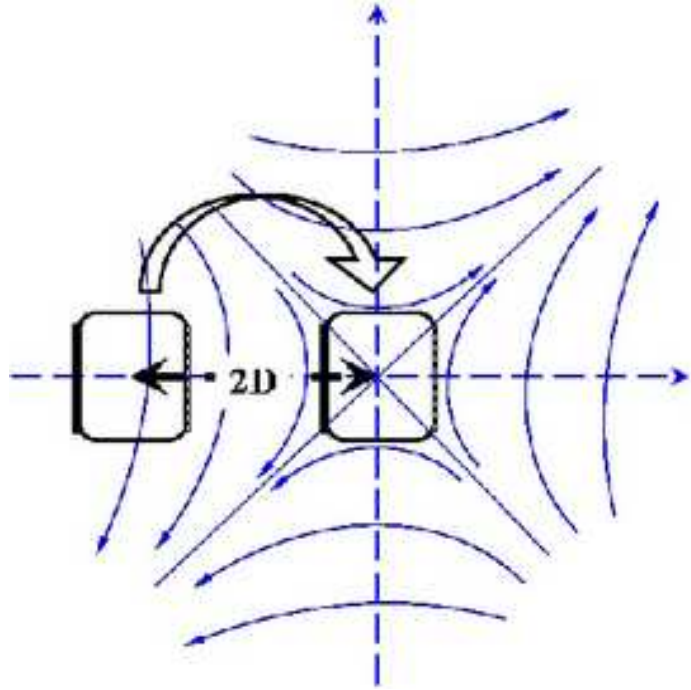} 
\caption[]{Coil set to measure the gradient and field quality of a
           quadrupole magnet}
\label{fig:quadcoil}
\end{figure}

The measurement of the strength of a quadrupole magnet could be done
by the displaced coil method of \Sref{sec:coils-displaced}. High
accuracy requires high precision displacement of the coils and the
measurement of the field quality, i.e., higher order multipoles, leads
to a uselessly complex mathematical treatment.  The array of two coils
connected in electrical opposition and sketched in \Fref{fig:quadcoil}
is often used to give the quadrupole strength by flipping them about
their symmetry axis. The coil array, \ie the distance $D$, can be
calibrated once in a known quadrupole. The field quality is deduced by
displacing, along the $x$ and $y$ axis, the array in a quadrupole in
the same way as the the single coil displacement to measure the field
quality of a dipole magnet
\Sref{sec:coils-displaced}). \Eref[b]{eq:coilarray1} is valid as long
as the two coils have an equal effective surface.

\begin{equation}
B_1(x+D) -  B_1(x-D) = 2 \cdot D \cdot G(x) ~. 
\label{eq:coilarray1}   \end{equation}

The idea of using two coils connected in opposition in order to cancel
the contribution of the dipole component is appreciably more
effectively used with the rotating coil method
(Sections~\ref{sec:dipcomp} and \ref{sec:quadcomp}). In that case, the
compensation scheme suppresses the main harmonic of the magnet. It is
therefore much more powerful for accurate field quality
measurements. The rotating coil equipment is more complex to
manufacture but is preferred for quadrupole magnets that are straight
and have a circular aspect ratio. The present method is, however,
mentioned as useful for combined-function magnets and pulsed magnet
measurements.

\subsection{The Morgan coil for pulsed magnets}

G. H. Morgan \cite{Morgan} proposed a complex coil array that can
measure any multipole magnet, or magnet component, in pulsed mode,
i.e., with static coil array. Identical coils are located around a
cylinder frame with the symmetry to be measured. The number of coils
is equal to the multipole order to be measured: three coils for a
sextupole, etc.

\Fref[b]{fig:BNLArray} details an array of coils able to measure a
large number of harmonics by different connection schemes to put in
series the individual coils. Small discrepancies between the
individual coils can be eliminated by turning the coil frame at
different angular positions. One should not forget that a coil scheme
sensitive to a multipole of order $n$ is as well sensitive to order
$n\cdot (2m+1)$. Reference~\cite{Jain2008} describes the full theory
of this technique.

\begin{figure}
\centering
\includegraphics[width=95mm]{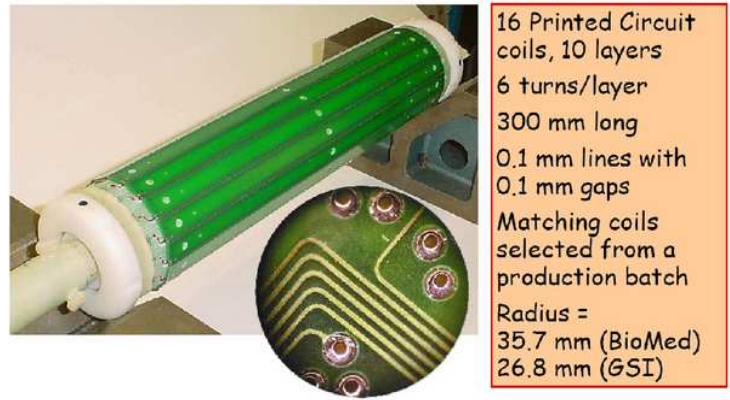} 
\caption[]{Measuring shaft with 16 individual coils made with printed
           circuit technology. A large number of harmonics can be
           measured in pulsed mode by using various connection schemes
           (courtesy of A. Jain, BNL).}.
\label{fig:BNLArray}
\end{figure}
 
\subsection{A dedicated coil array to measure curved dipoles in pulsed mode: 
            the CNAO fluxmeter}
\label{sec:CNAO}

A dedicated array of 11 curved coils has been assembled to measure the
CNAO dipole magnets \cite{Rossi2006}. These 1.5 T dipoles have a
bending of $22.5^{\degree}$ and operate with a field rise time of
2~seconds. Eddy current effects present in particular in the ends of
the magnet yoke have to be taken into account for both the field
integral as a function of current and the field quality over the
130~mm useful aperture.

\begin{figure}
\centering
\includegraphics[width=95mm]{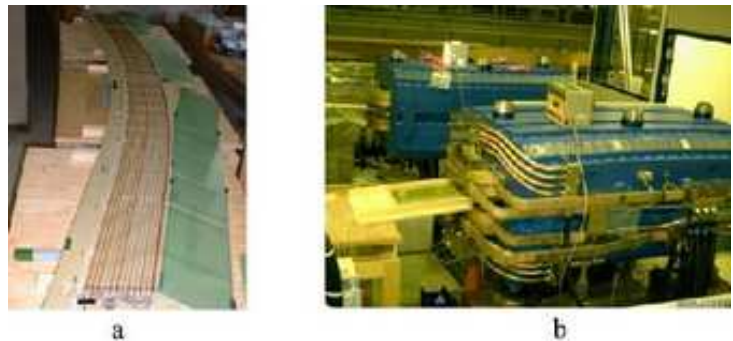} 
\caption[]{(a) The CNAO fluxmeter with 11 curved coils, (b) inserted
           in one of the CNAO dipole magnets}
\label{fig:cnaoarray}
\end{figure}

\Fref[b]{fig:cnaoarray}(a) shows the 11 coils fixed on the frame
that can be entered from a zero field region into one magnet end to
measure the remanent field, or measure in static position during the
field ramp [\Fref{fig:cnaoarray}(b)]. A cross-calibration 12th coil
can be mounted on top of each individual coil of the fluxmeter. This
cross-calibration gave correction factors to equalize the effective
width of the 11 coils within $10^{-4}$ relative accuracy.

Note that all correction factors described in \Sref{sec:realcoil} have
to be applied to calculate the field errors in terms of
multipoles. \Fref[b]{fig:cnaocurves} gives the vertical field
component in the horizontal symmetry plane of 10 CNAO dipole magnets
measured with this fluxmeter, taking into account these systematic
errors.

% SSWdipole.tex

\section{The single stretched wire technique with a dipole magnet}
\label{sec:sswdipole}

\begin{figure}
\centering
\includegraphics[width=105mm]{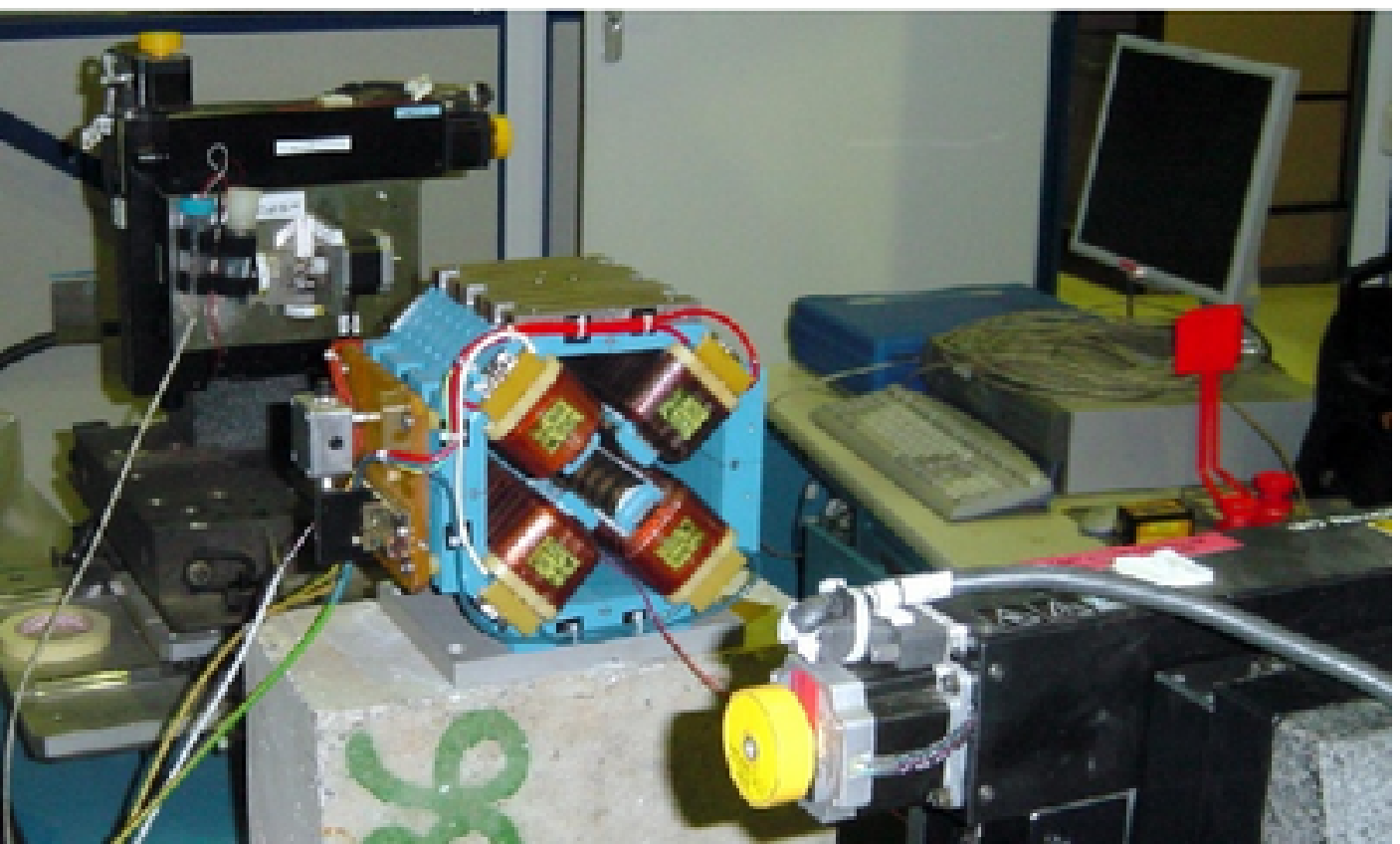} 
\caption[]{The SSW equipment installed to measure a quadrupole
           magnet. The wire is displaced in the quadrupole aperture by
           2D high-precision stages on both sides.}
\label{fig:ssw}
\medskip
\centering
\includegraphics[width=105mm]{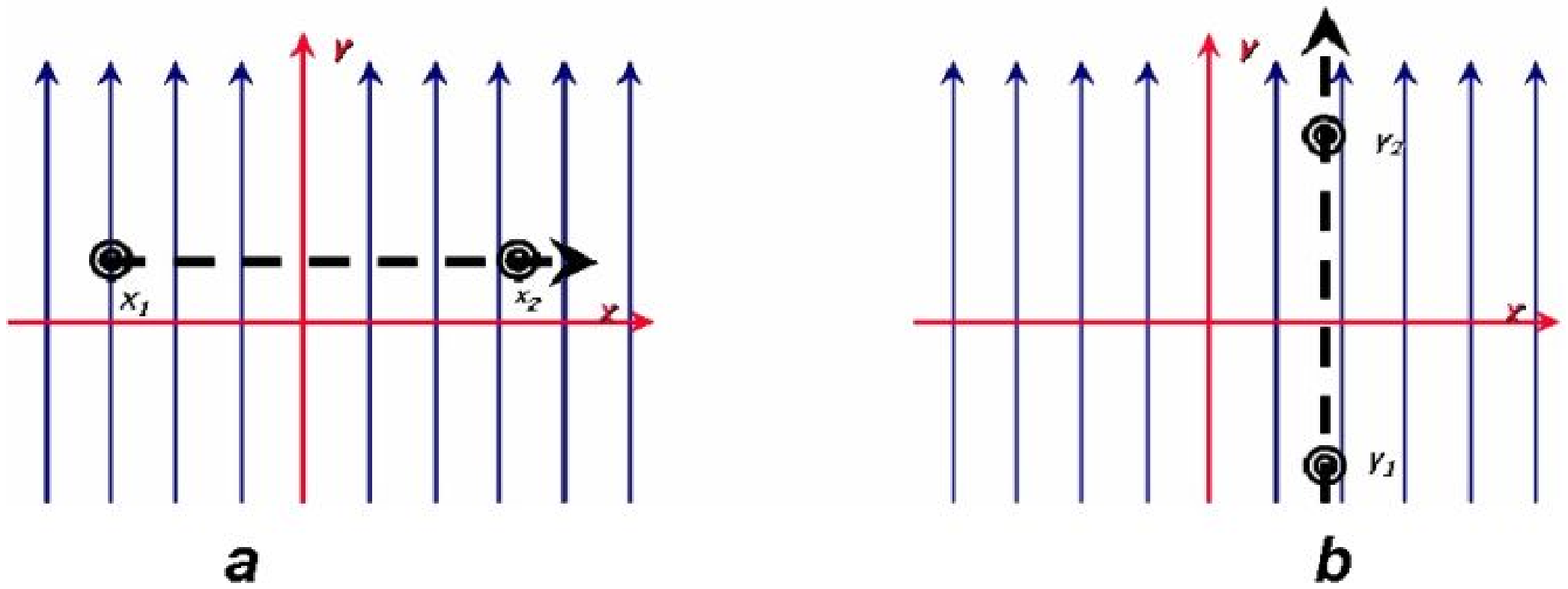} 
\caption[]{(a) The single stretched wire displaced horizontally to
           measure the field integrated over the dipole length. (b) A
           vertical displacement gives zero signal if displacement and
           field direction are parallel.}
\label{fig:sswdipole}
\end{figure}

Particle beams are sensitive to field integrated along their
trajectory. The SSW (Single Stretched Wire) method consists of a high
tensile conducting wire moved inside the magnet aperture by precision
displacement tables. CuBe wires 0.1\Umm{} thick are commonly used. The
stages at both sides are assumed to move by precisely the same
amount. The return wire is kept fixed, as much as possible in a
field-free region. The flux lines crossed during this displacement,
$\Psi(x_1,x_2)$, and measured by a voltage integrator give the field
integrated over the displacement, $d=x_2 - x_1$, and over the SSW
length, $L_\text{w}$. \Fref[b]{fig:ssw} shows the system developed by
FNAL \cite{Dimarco2000} and mounted to measure a quadrupole
magnet. When measuring a perfect dipole ($B_y = Cst~~ , ~~B_x = 0$) as
in \Fref{fig:sswdipole}(a) the integrator gives
\begin{equation}
\Psi(x_1,x_2)= \int_{0}^{L}{\int_{x_1}^{x_2}{B_y(x,l)} \cdot dx \cdot dl}
             = d \cdot \int{B \cdot dl}~.
\label{eq:ssw1}   
\end{equation}

The measurement accuracy of the dipole strength is linked to the
calibration of the integrator gain $(10^{-4}- 10^{-5})$ and to the
precision of the mechanical displacement. Commercially available
stages reach an accuracy better than $10^{-4} =
1\Uum\text{/}10\Umm$. This method was cross-checked 30 years ago
against NMR mapping and gave agreement within few $10^{-5}$.

Similarly, a vertical displacement in a dipole aligned vertically
gives a zero flux variation [see \Fref{fig:sswdipole}(b)]. It is the
simplest and most accurate method of finding the field direction of a
dipole: accuracies of 0.1~mrad are commonly reached.

This method is simple and efficient to measure the first integral of
wigglers and undulators. This first integral value, given by
\Eref{eq:ssw1}, corresponds to the angular deflection of the beam and
is tuned to be zero in the relevant cross-section of the magnet. It
has been complemented by the pulsed wire technique to measure and tune
the second field integral value~\cite{Ruland1999,Fan2002}. Travelling
Hall plate based measurements are nowadays preferred since in addition
they give more details on the regularity of the undulator periods.
% SSWaxis.tex

\section{Align a quadrupole with the single stretched wire}
\label{sec:sswaxis}

The single stretched wire technique is relevant to finding the axis,
main field direction, and longitudinal position of a quadrupole
magnet~\cite{Dimarco2000}. These three alignment steps will be treated
separately so as to be easily understood. An automated acquisition
system is better for iterating changes in the reference position of
the wire for both stages according to full measurement cycles until an
accurate coincidence between wire and magnet axes is reached.

A pure quadrupolar field is defined by 

\begin{equation}
B_y = G \cdot x\quad,\quad B_x = G \cdot y~.
\label{eq:ssw2}   
\end{equation}

The magnetic axis is defined by the line where
\begin{equation}
B_x =B_y = 0~.
\label{eq:ssw22}   
\end{equation}

The main field direction is defined by the symmetry planes: 
\begin{itemize}
\item $B_x=0$ in the horizontal symmetry plane, 
\item $B_y=0$ in the vertical symmetry plane. 
\end{itemize}

Moving a SSW vertically from position $y_1$ to position $y_2$
[\Fref{fig:sswquadrupole}(a)] gives the measured flux of
\Eref{eq:ssw3}, \ie a parabolic dependence. The effective
length $L_\text{eff}$ hides the integral over the magnet length.

\begin{equation}
\Psi(y_1,y_2)=\int_{0}^{L}{\int_{y_1}^{y_2}{G \cdot y \cdot dy \cdot dl}}
             = L_\text{eff} \cdot \frac{G}{2} \cdot (y_2^2-y_1^2)~. 
\label{eq:ssw3}   
\end{equation}

\begin{figure}
\centering
\includegraphics[width=\linewidth]{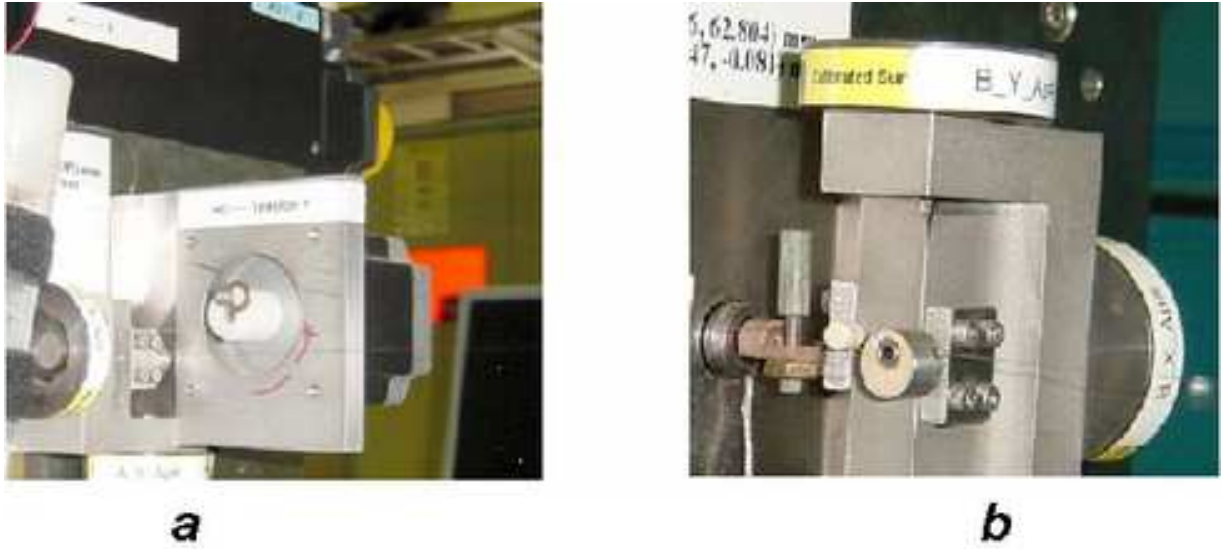} 
\caption[]{(a) Stepping motor to adjust the wire tension in order to
           extrapolate the measurement to infinite tension. (b)
           Tension gauge used to drive the stepping motor at the
           expected tension.}
\label{fig:tension}
\end{figure}

A further correction must be applied to take into account the wire
sagitta that could reach millimetres for a 10 to 15~m long distance
between the stages. The system of \Fref{fig:ssw} incorporates a driver
to make the measurement with different wire tensions
(\Fref{fig:tension}). The measurement data are extrapolated to
infinite tension. This will be detailed in \Sref{sec:extrapolate}
since this error source is more detrimental for the measurement of the
strength of a quadrupole. To obtain accuracy in the result requires
therefore time, even with fully automated equipment and procedures,
since loops at different tensions are internal to iteration to align
the wire coordinate system to the magnet axis. An accurate measurement
of the gradient can only take place after a full alignment procedure.

\subsection{Align the quadrupole axis and field direction}
\label{sec:sswquadaxis}

\begin{figure}
\centering
\includegraphics[width=125mm]{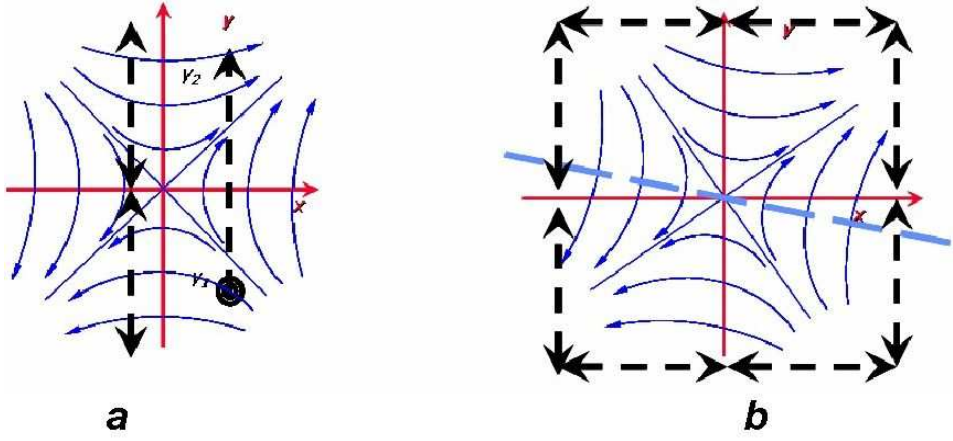} 
\caption[]{(a) The single stretched wire displaced vertically in a
           quadrupole. Two measurements are needed to find the minimum
           of the parabolic dependency, i.e., the horizontal symmetry
           plane. (b) The axis and field direction (i.e., tilt of the
           quadrupole field) are found with eight measurements.}
\label{fig:sswquadrupole}
\end{figure}

\Eref[b]{eq:ssw3} tells us that two measurements are needed to find
$y_c$, minimum of the parabola giving the horizontal plane where the
axis is located. The method in use is to measure in an iterative way
the fluxes over two equal intervals then displace the central point of
the measurement $y_c$ until the following condition is reached:

\begin{align}
\Psi(y_c,y_c+d)&=\Psi(y_c,y_c-d)~,
\label{eq:ssw4} \\
(y_c+d)^2- y_c^2 & = (y_c-d)^2- y_c^2~.
\label{eq:ssw5}   
\end{align}

The point where $B_x=0$ is found for a given vertical line. The main
field direction is found by displacing this movement line along $x$,
and the vertical symmetry plane is found by equivalent horizontal
displacements. A full measurement is done by the eight displacements
of the external square of \Fref{fig:sswquadrupole}(b) and repeated
until the SSW coordinate system coincides with the quadrupole axis. It
is obviously important that the two stages move parallel to each
other.

\subsection{Stretched wire non-parallel to the quadrupole axis}
\Fref[b]{fig:sswalign}(a) sketches, for the vertical direction,
how to tune the parallelism between the SSW and magnet axes. A
displacement of one stage going from an angle $\alpha_1$ to $\alpha_2$
with respect to the magnet axis gives a flux:

\begin{equation}
\Psi(\alpha_1,\alpha_2)
  =\int_{\alpha_1}^{\alpha_2}{\int_{L_1}^{L_2}{G
         \cdot \alpha \cdot l \cdot dl \cdot d\alpha}} 
~~=~~ G \cdot \frac{\alpha_2^2-\alpha_1^2}{2} \cdot \frac{L_2^2-L_1^2}{2}~.
\label{eq:ssw6}   
\end{equation}
The effective length of the magnet $(L_2-L_1)$ should be known, and
the middle of the magnet is supposed to coincide with the middle of
the wire. This is, however, of minor importance since this method,
similarly to the one of \Sref{sec:sswquadaxis}, implies a series of
two symmetric displacements iterated until the parallelism is found,
\ie until they give equal flux values.

\subsection{Measure the longitudinal location of the magnet}

\begin{figure}
\centering
\includegraphics[width=\linewidth]{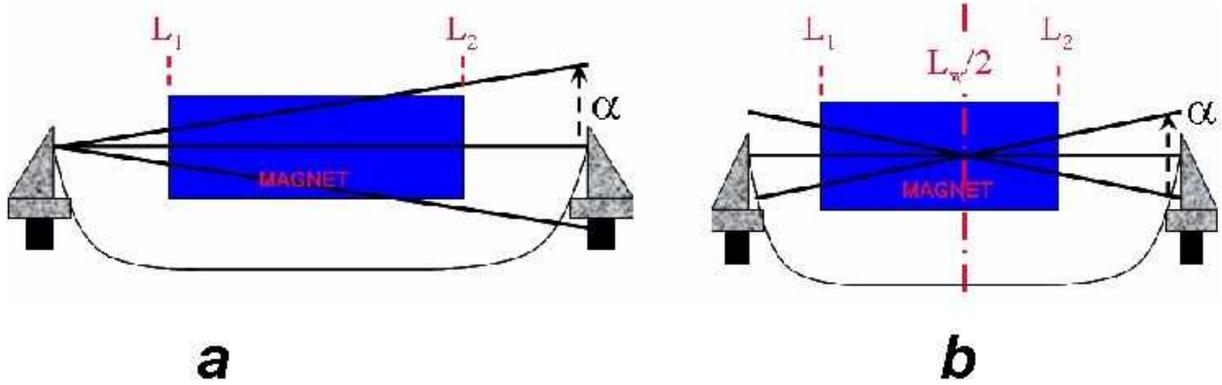} 
\caption{(a) Two angular displacements give the same result if done
         symmetrically with respect to the quadrupole axis, \ie when
         this axis and the reference line of the SSW coincide. (b) The
         same angular displacements with the fixed point at the middle
         of the SSW length give the distance between this SSW middle
         and the middle of the magnet length.}
\label{fig:sswalign}
\end{figure}

Once the magnet is fully centred and the axis is parallel to the wire
reference position, anti-parallel movements of the wire give the
longitudinal position $d$ of the magnet with respect to the middle of
the wire length $L_\text{w}$ [\Fref{fig:sswalign}(b)]. The angular movement
$\alpha$ has to be symmetric with respect to the magnet axis. The flux
for such angular movement is given by

\begin{equation}
\Psi(\alpha_1,\alpha_2) 
 = \int_{\alpha_1}^{\alpha_2} \int_{L_1}^{L_2}
    G \cdot x(\alpha \cdot l)\cdot dl \cdot d\alpha  
 = G \cdot \frac{\alpha_2^2-\alpha_1^2}{2} \cdot d \cdot L_\text{eff}~,
\label{eq:ssw7}   \end{equation}
 
with 
\begin{gather*}
  x   =\left(l-\frac{L_\text{w}}{2}\right) \cdot \alpha ~,\quad
  L_1 = \frac{L_\text{w}-L_\text{eff}}{2}+d ~,\quad
  L_2 = \frac{L_\text{w}+L_\text{eff}}{2}+d~. 
\end{gather*}
  
\subsection{Automatic alignment of the SSW reference system to the quadrupole 
            axis in all directions}

The system has to be designed to automatically align the SSW with
respect to a quadrupole according to these various principles put
together. The SSW reference system is aligned iteratively, by one
compound measurement with eight co-parallel and eight counter-parallel
displacements according to \Fref{fig:sswquadrupole}(b), followed by
modifications of the SSW reference system. This alignment campaign is
tedious since each of these 16~data sets has to be done with different
tensions of the wire to extrapolate the results to infinite tension,
as detailed in \Sref{sec:extrapolate}. The longitudinal location is
calculated with the last measurement data.  The precision obtained
with the FNAL equipment for wires up to 16\Um{} long is remarkable:
about 0.1\Umm{} for the axis at both magnet ends and 0.03\Umm{} for
the integral, 5\Umm{} for the longitudinal position.
 
% SSWquadrupole.tex

\section{Measure the integrated strength of a quadrupole with the SSW}
\label{sec:sswquadrupole}

The difficulties encountered when measuring a quadrupole strength are
that the wire has a natural deflection in the millimetre range if the
wire or magnet length reaches several metres. In addition, it is
difficult to find CuBe wire that has zero magnetic
susceptibility. Unfortunately the industrial standards describing this
type of material rarely include the impurity content of non-zero
susceptibility. Several batches purchased from the same manufacturer
could have appreciably different magnetization and the only way to
sort the best batch is to test with a permanent magnet.  In conclusion
the wire deflection depends on the position in the quadrupole
cross-section, in both amplitude and direction. The accuracy is also
limited by the high order multipoles present in the magnet that
perturb the value of $G \cdot L_\text{eff}$ obtained from
Eqs.~(\ref{eq:ssw3})--(\ref{eq:ssw5}). The only way to estimate
accurately that error source is to perform a full measurement of the
multipole content for instance with the rotating coil system
(\Sref{sec:Harmcoil}). These perturbations grow with the
distance from the magnet axis, and a detailed estimation is needed to
either limit the range of the displacement allowed or to include
relevant correction factors in the data analysis.  In practice, the
estimation of the quadrupole strength is obtained from the last set of
data measured once the SSW reference system is fully aligned with the
quadrupole axis (\Sref{sec:sswaxis}).

\subsection{Extrapolate the results to infinite tension of the wire}
\label{sec:extrapolate}

The sagitta due to gravity can easily amount to millimetres with a
CuBe wire at maximum tension. The wire tension is measured by the
vibration frequency $f$ to get enough precision. This vibration can be
induced by the stage motors giving a horizontal or vertical kick and
is sensed via the voltage across the wire since it oscillates in a
magnetic field. An extrapolation to infinite tension is usually done
by looking at the curve of the integrated gradient measured as a
function of the measured $1/f^2$, i.e., proportional to the wire
tension according to the following relations:

\begin{equation}
\text{sag} =\frac{W\cdot g}{8T}L_\text{w}^2
\label{eq:sag}
\end{equation}

\begin{equation}
f=\frac{1}{2L_\text{w}}\sqrt{\frac{T}{W}}
\label{eq:vibrationfrequency}
\end{equation}
with $W$ the weight of the wire per unit length, $g$ the gravity
~constant, and $T$ the wire tension.

\Fref[b]{fig:tension}(a) shows the remotely controlled actuator to
change the wire tension. \Fref[b]{fig:tension}(b) shows a tension
gauge used to control the adjustment of the tension in particular to
avoid breaking the wire.

\subsection{Measure the gradient with horizontal displacements of the wire}
\label{sec:measure gradient} 

\Fref[b]{fig:frequencydependance} shows the curves of the transfer
function obtained at different currents into an LHC quadrupole
magnet~\cite{smirnov2006} for a given type of CuBe stretched
wire. Three observations can be made about this measurement campaign
summarized in \Tref{table:gradientslope}:
\begin{itemize}
\item the slope of the measured gradient depends, as expected, on the
      frequency but also on the square of the gradient, \ie the
      current in the quadrupole magnet;
\item gradient values measured by vertical displacements were found to
      be systematically higher than with horizontal displacements;
\item this dependence on the gradient is different according to the
      batch of CuBe wire used, and this finding correlates with the
      wire being paramagnetic or diamagnetic as sensed by a permanent
      magnet.
\end{itemize}

\begin{figure}
\centering
\includegraphics[width=8cm]{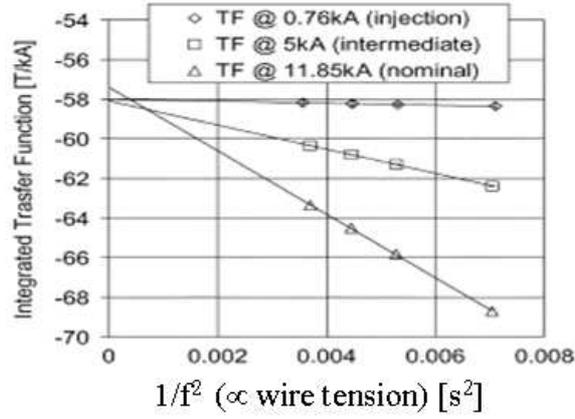} 
\caption[]{Measured gradient, integrated over the quadrupole length,
           divided by the current as a function of the wire tension
           measured by the vibration frequency}
\label{fig:frequencydependance}
\end{figure}

\begin{table}
\caption[]{Slopes of measured quadrupole strength, as a function of
           inverse squared vibrating frequency, for different types of
           wire, in [$\UTZ/\UsZ^2$].}
\label{table:gradientslope}
\centering
\begin{tabular}{@{}crrrl@{}}                                \hline \hline
Wire No. & 0.76 kA & 5 kA & 11.5 kA & $\chi$              \\\hline
1        & 30.4~~  & 2000 &  9500~~ & paramagnetic        \\
2        &  6.1~~  &  500 &  5000~~ & paramagnetic        \\
3        &  2.3~~  &   50 &   474~~ & diamagnetic         \\
4        & --~~    &  --~ &   380~~ & diamagnetic         \\\hline \hline
\end{tabular}
\end{table}

A detailed analysis proved that the $1/f^2$ dependence was linear for
horizontal displacements but not for vertical displacements. This is
explained by looking at the dependence of the magnetic forces in both
cases.  The force variation due to the field gradient is horizontal
for an horizontal displacement:
\begin{equation}
  F_x \propto G^2 \cdot(x_\text{step})^2~,
\end{equation}
but has a parabolic dependence with vertical displacements
\begin{equation}
  F_y \propto G^2 \cdot(y_\text{step}+ \text{sag})^2~.
\end{equation}

The gradient obtained with vertical displacements should therefore be
calculated with a fit to a parabolic law, so is less stable than for
the horizontal displacements. It is therefore better to choose the
result obtained with the horizontal displacements.

Relative discrepancies of $2 \cdot 10^{-4}$ for the measured gradient
were obtained between measurement with vertical and horizontal
displacements, giving an estimate of the accuracy that can be obtained
with this method.
% Vibrating.tex

\section{The vibrating wire technique}
\label{sec:Vibrating}

\begin{figure}
\centering
\includegraphics[width=12cm]{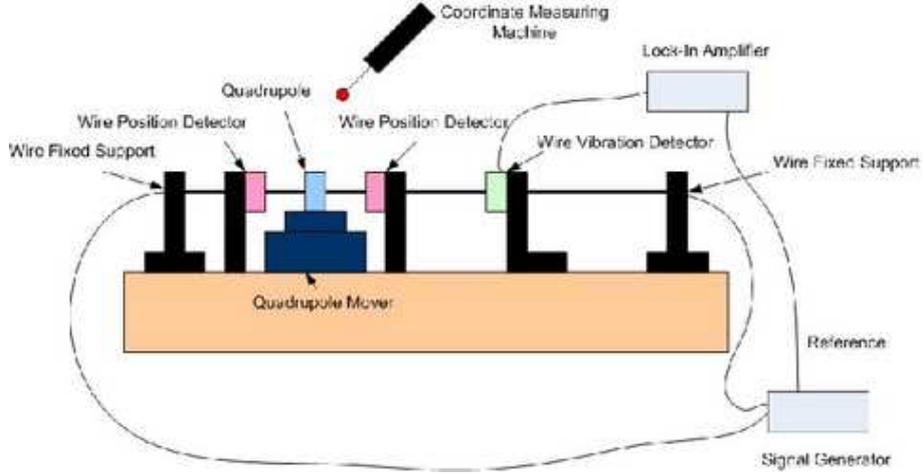} 
\caption[]{Overview of the vibrating wire equipment measuring the
           magnetic axis in a quadrupole. The wire vibration detector
           and the quadrupole are longitudinally located to have
           maximum signal with the second harmonic of the natural
           oscillation frequency (courtesy of Z. Wolf, SLAC).}
\label{fig:vibratingoverview}
\end{figure}

Mechanical oscillations of the stretched wire can be induced by the
Lorentz force created by AC current flowing into the wire going
through the static field of the measured magnet. This technique,
proposed by A. Temnykh, has sub-micrometre sensitivity to sense a
quadrupole axis. It can be used to measure separately the axis of
several magnets aligned on a girder and has been extended to find
sextupole axes.

\Bref[b]{Temnykh1997} details the resolution of the differential
equation to solve for standing wave solutions:

\begin{equation}
W \frac{\partial x^2}{\partial t^2}=
T\frac{\partial x^2}{\partial z^2}- \gamma \frac{\partial x}{\partial t} + I(t)B(z)
\label{eq:vb1}
\end{equation}
with $x(0,t) = x(L_\text{w},t) = 0$ the boundary~conditions, $W$ the
weight of the wire per unit length, $T$ the wire~tension, $\gamma$ the
damping~coefficient, $I(t) = I_0 \cdot e^{i\omega t}$
the~driving~AC~current, and $B(z)$ the transverse~magnetic~field.
  
The solution is a sum of standing waves of amplitudes $x_n$ that can
be measured. From there the $B_n$ coefficients of a sinus wave
expansion of $B(z)$ are estimated
\begin{equation}
x(z,t) = \sum{x_n \sin\left(\frac{\pi nz}{L_\text{w}}\right)e^{i \omega t}}~,
\label{eq:vb2}
\end{equation}
with
\begin{align*}
 x_n    
 &= \frac{I_0}{W} \cdot \frac{1}{\omega^2 - \omega_n^2 - i \gamma \omega}\cdot B_n~,\\
\omega_n 
 &= \frac{\pi n}{L_\text{w}}\sqrt{\frac{T}{W}}~.
\end{align*}
  
$B(z)$ can be reconstructed knowing the series of $B_n$:
\begin{equation}
B(z) = \sum{B_n \sin\left(\frac{\pi nz}{L_\text{w}w}\right)~.}
\label{eq:vb3}
\end{equation}

In practice, the AC current is tuned to the natural oscillation
frequency $f_\text{nat}$ of the wire or one of its harmonics $m \cdot
f_\text{nat}$.  The amplitudes of these oscillations are measured by
inexpensive photo interrupters.  The wire motion in the vertical,
respectively horizontal, plane is caused by the Lorentz forces between
the wire current and the horizontal, respectively vertical, magnetic
field. Therefore two wire vibration detectors are mounted orthogonal
to each other.

\begin{figure}
\centering
\includegraphics[width=12cm]{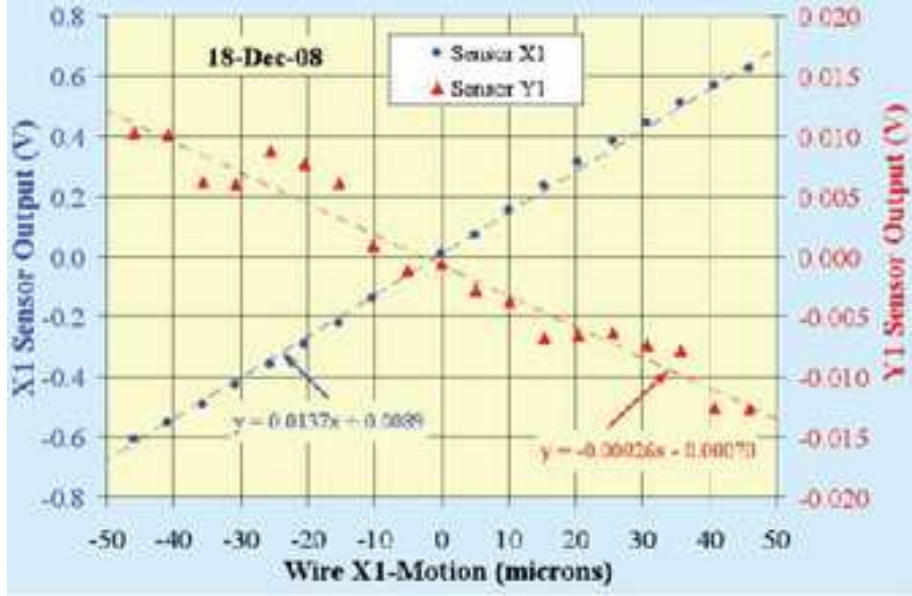} 
\caption[]{Amplitudes measured by the $x$ and $y$ photocouplers for
           small displacements of the vibrating wire around the
           quadrupole axis in the $x$ direction (courtesy of A. Jain,
           BNL)}
\label{fig:vbcoupling}
\end{figure}

A procedure to centre one quadrupole is straightforward. The wire
oscillation detectors can be located in a longitudinal position where
the amplitude is large for the vibration harmonic considered. That is
why it is separated from the position detectors in
\Fref{fig:vibratingoverview} that gives an overview of the
equipment~\cite{Wolf2005}. Since the Lorentz force is zero if the wire
is aligned with the quadrupole axis, the magnet or the wire can be
moved until vanishing oscillations are seen in both planes. Some
coupling between the measured oscillations in both directions could
exist due to imperfect orthogonality between the detectors. It has to
be measured and taken into account to achieve high accuracy
(\Fref{fig:vbcoupling}). These measurements obtained by wire motion
controlled by piezo-electric actuators demonstrate the sub-micron
resolution~\cite{Jain2009}.

% Harmcoil

\section{Description of the harmonic coil method}  
\label{sec:Harmcoil}

The harmonic coil method was developed with early analog integrators
forcing the measuring coil to rotate stepwise between consecutive
angular positions~\cite{Cobb1970}. Fast angular encoders and
purpose-developed voltage integrators with zero dead time between the
angular positions are the basis of today's systems acquiring several
hundred points per turn with a rotation rate as high as 10 turns per
second. It is the best method for measuring higher order multipoles
within a well-established theoretical frame, in particular of
superconducting and quadrupole magnets having circular
apertures. Progress in data acquisition equipment and data analysis
tools alleviates the complexity of the formalism applied to the amount
of data to treat, so that fully automated instruments and data
analysis processes have been developed for measurements of series of
magnets with high confidence in the final results.

\subsection{Flux enclosed by a simple rotating coil}

\begin{figure}
\centering
\includegraphics[width=13cm]{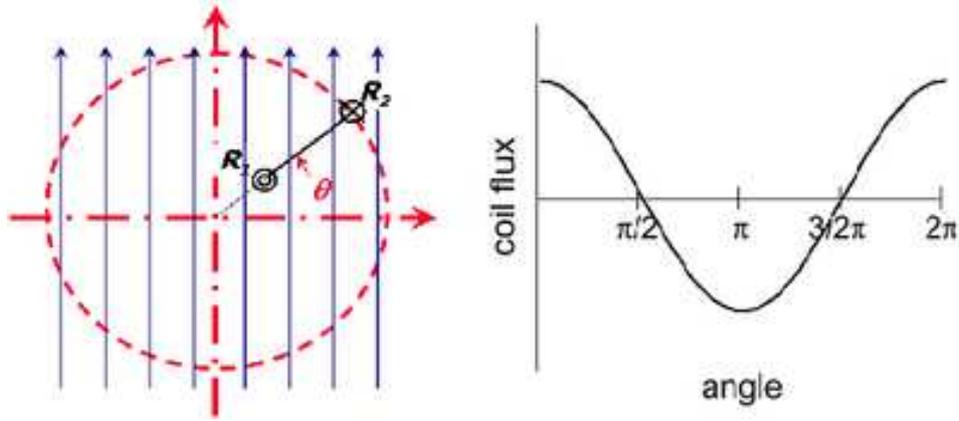} 
\caption[]{2D representation of the flux seen by a simple coil
           rotating in a dipole field}
\label{fig:simplecoil}
\end{figure}

A perfect dipole magnet gives a constant vertical field everywhere in
the useful aperture. The flux enclosed by the simple coil described in
\Fref{fig:simplecoil} will be, considering an infinitely thin winding,

\begin{equation}
\Psi(\theta)= N_\text{t}\cdot L \cdot \int_{R_1}^{R_2}{B_1 \cdot \cos\theta \cdot dR} ~. 
\label{eq:eq1}   \end{equation}

$N_\text{t}$ and $L$ are respectively the number of turns and length of the
measuring coil. The coil is supposed to be shorter than the
magnet. The coil's effective surface can be calibrated independently
and is given by

\begin{equation}
\Sigma_\text{coil} = N_\text{t}\cdot L \cdot \int_{R_1}^{R_2} dR 
                   = N_\text{t}\cdot L \cdot(R_2-R_1)~.
\label{eq:eq3}   
\end{equation}

The use of a voltage integrator connected to the measuring coil makes
it possible to eliminate the time coordinate in the induction law of
Faraday. The voltage integrator read as a function of the angle gives
the flux directly from the zero angle where it is reset. The constant
of integration is irrelevant for this method.

The units of \Eref{eq:eq1} are
\begin{equation}
\Psi  = \text{volt} \cdot \text{sec}  
      = \text{tesla} \cdot \text{m}^2  
      = \text{weber}~. 
\label{eq:eq2} \end{equation}

It is important to realize that the harmonic coil method does not make
use of the voltage integrated over a given time, but rather over a
given angular interval. The advantage of using a voltage integrator
that can be externally triggered is that it eliminates to the first
order the problem of a constant speed of rotation.  A real system in
fact measures differences of fluxes between two~incremental angular
positions. The angular encoder mounted on one coil end is a
fundamental piece of equipment, as described in
\Sref{sec:encoder}. The integrator is triggered by this encoder
and collects incremental fluxes $\delta\Psi_k,$ and the left part of
\Eref{eq:eq1} becomes
\begin{equation}
\Psi(\theta_i) - \Psi(\theta_{0})= \sum_{k=1}^i{\delta\Psi_k}~,   
\label{eq:eq5}
\end{equation}
with 
\begin{equation*}
\delta\Psi_k  = \Psi(\theta_k) - \Psi(\theta_{k-1})~. 
\label{eq:eq6} 
\end{equation*}

\subsection{Formalism for multipoles measured by rotating coils}
\label{sec:basic} 
The power of the harmonic coil method is its ability to measure any
type of 2D magnetic field. It can be demonstrated~\cite{Davies92} that
a rotating coil measures the 2D field integrated over its length as
long as the field component parallel to the rotation axis is zero on
the two coil ends. The complex equation to best describe this 2D field
is
\begin{equation}
B(x+i\cdot y)= B_y(z) + i \cdot B_x(z) 
             = \sum_{n=1}^{\infty} {C_n \cdot \left(\frac{z}{R_\text{r}}\right)^{n-1}} ~.
\label{eq:eqH6}  \end{equation}

The components $C_n=B_n + iA_n$ are the normal and skew multipoles of
the field. By definition for accelerator magnets, the normal
components indicate a vertical field in the horizontal plane whilst
the `skew' terms apply for an horizontal field. The $C_n$ are in tesla
at the reference radius $R_\text{r}$. \Fref[b]{fig:mainB} shows the
field lines for normal and skew dipoles $(C_1)$ and quadrupoles
$(C_2)$.

\begin{figure}    
\centering
\includegraphics[width=12cm]{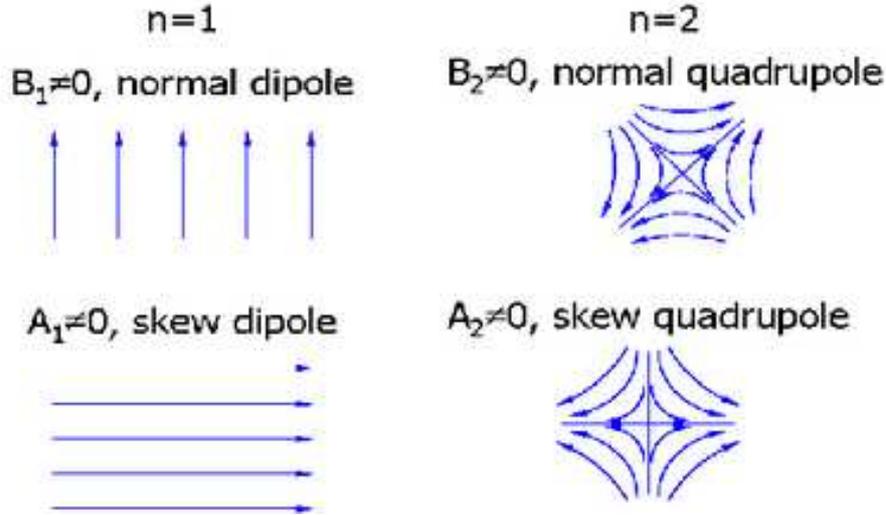} 
\caption[]{Field lines of normal and skew dipole and quadrupole magnets}
\label{fig:mainB}
\end{figure}

The field quality is usually described as errors relative to the main
field component $B_M$ ($M=1$ for a dipole, $M=2$ for a quadrupole) at
the reference radius $R_\text{r}$. These errors are called `units' and
are given by
\begin{equation}
c_n = b_n + i \cdot a_n = 10^4 \frac{C_n}{B_M} ~.
\label{eq:eqH7}  
\end{equation} %units fo 10-4

The reference radius $R_\text{r}$ is an important concept for
accelerator magnets having apertures much smaller than one
metre. $R_\text{r}$ corresponds in practice to
\begin{itemize}
\item the useful aperture for the beam,
\item 2/3 of the yoke aperture in resistive magnets,
\item 2/3 of the coil aperture in superconducting magnets,
\item the radius where the multipoles relative to the main field, the
      $c_n$ in \Eref{eq:eqH7}, have the same order of magnitude in a
      real magnet.
\end{itemize}

It is important to carefully choose this reference radius at the
beginning of a project. It will intervene in the discussions between
all actors involved: beam optic physicists, magnet designers,
measurement crew, and data analysis teams.

The voltage integrated over a simple rotating coil described in
\Fref{fig:simplecoil} and rotating in any 2D field will therefore be
\begin{equation}
\Psi(z)= N_\text{t}\cdot L \cdot \text{Re}\int_{R_1}^{R_2}{B(z)\cdot \rmd z}~.
\label{eq:eqH8}   \end{equation}
Since the coil rotates
\begin{equation}
z = x+i\cdot y=R\cdot e^{i\theta(t)}~.
\label{eq:eqH9}  \end{equation}
And by applying \Eref{eq:eqH6} and integrating it over $\rmd R$
\begin{equation}
\Psi(\theta=\omega\cdot t)
           =\text{Re}\left(\sum_{n=1}^\infty N_\text{t}\cdot L \cdot 
                           \frac{R_2^n-R_1^n}{n\cdot R_\text{r}^{n-1}} \cdot 
                           C_n\cdot e^{in\theta}
                     \right) ~.
\label{eq:eqH10}  
\end{equation}
This allows a formal separation between what belongs to
\begin{itemize}
\item the measured field components $C_n$,
\item the time dependence of the signal $e^{\rmi n\theta(t)}$,
\item the coil sensitivity factor $K_n$ defined as
\end{itemize}

\begin{equation}
K_n=N_\text{t}\cdot L \cdot \frac{R_2^n-R_1^n}{n\cdot R_\text{r}^{n-1}}~.
\label{eq:eqH11}   
\end{equation}
 
The $K_n$ are calculated once for each measuring coil used. As will be
detailed in \Sref{sec:coil} they can be complex numbers in the
case of tangential coils, or coils not perfectly aligned
radially. These calculations are substantial if the windings can no
longer be considered point-like. Their values can be improved by
individual calibrations~\cite{CAS09MarcoB}.

The power of the harmonic method is that the multipoles of the field
are directly given by the Fourier analysis coefficients $\Psi_n$ of
the integrated voltage over a coil turn $\Psi(\theta)$:

\begin{equation}
\Psi_n=K_n \cdot C_n=K_n \cdot (B_n+iA_n)~.
\label{eq:eqH12}   
\end{equation}
 
\subsection{Centre the measurement results by measuring the field axis and direction}

\begin{figure}[ht]  
\centering
\includegraphics[width=12cm]{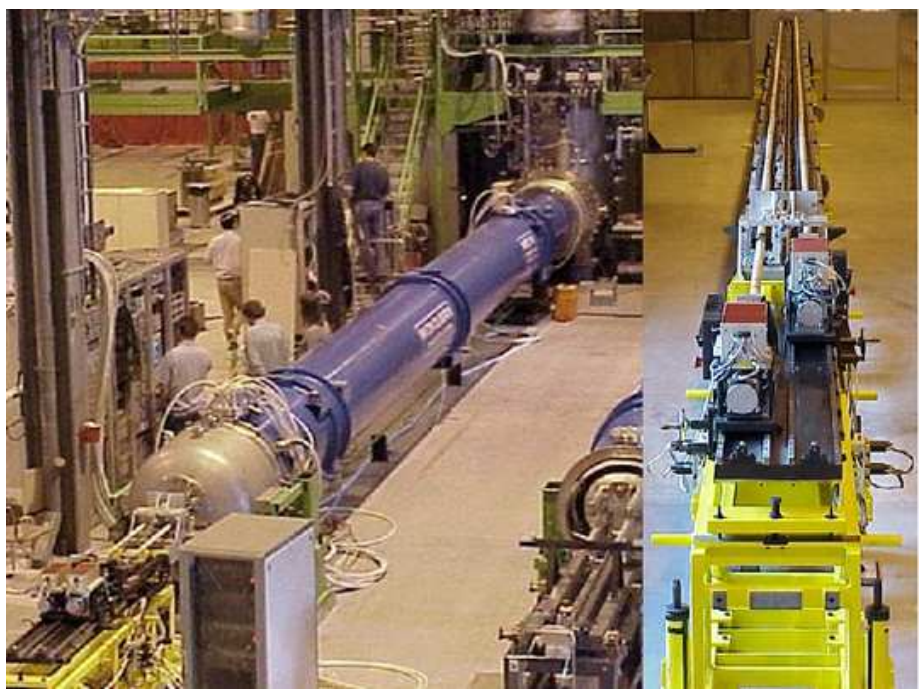} 
\caption[]{LHC superconducting magnets measured with a 15\Um{} long
           shaft. The `Twin Rotating Unit' at the bottom of both
           pictures holds the motors and encoders.}
\label{fig:SM18}
\end{figure}

The harmonic method measures with high accuracy the direction of the
main field component with respect to the measuring coil direction. It
gives as well the axis of a quadrupole or sextupole magnet with
respect to the rotation axis of the measuring coil. These measurements
are fundamental for accelerator magnets since the beam closed orbit is
defined by the residual tilt of the dipole magnets and the quadrupole
misalignments. This property is also useful to align the set of
measured multipoles in the symmetry axis of the magnet. This was
required, for instance, for the system described in \Fref{fig:SM18}
where the 12 intermediate bearings of the 15 m long measuring shafts
are poorly centred with respect to the superconducting magnet. Simple
equations given below formally change the reference axis system of the
multipole set.

\subsubsection{Find the main field direction}
\label{sec:Field direction}

The harmonic coil method gives the field description, the $C_n$ in
\Eref{eq:eqH12}, in the reference frame defined by the measuring
coil when at $\theta$ = 0, usually given by the reference angle of the
encoder. The main field direction corresponds to a zero main skew term
($A_1$ = 0 for a dipole magnet). This main field direction often does
not correspond with enough accuracy to the magnet mechanical symmetry
plane, hence the usefulness of this measurement. In addition, both the
magnet and measuring system may not be aligned with gravity, used as
an external reference. Three frames are considered:

\begin{itemize}
\item $\theta_m$ : zero of the encoder, reference for the Fourier analysis;
\item $\theta_g$ : gravity, magnet fiducials when aligned;
\item $\theta_f$ : field direction of the magnet ($A_1$ = 0 for a dipole magnet).
\end{itemize}

To rotate the multipole coefficients from one frame to another is
straightforward from \Eref{eq:eqH10}. Assuming respectively
$\theta_m$ and $\theta_f$ the angles of the measurement frame and the
magnet symmetry plane with respect to gravity give:
\begin{align}
z_f    & =z_m \cdot e^{i(\theta_m-\theta_f)}~,
\label{eq:eqH13} \\
C_{m,n}& =C_{f,n} \cdot e^{in(\theta_m-\theta_f)}~.
\label{eq:eqH14}   
\end{align}

\begin{figure}[ht]    
\centering
\includegraphics[width=10cm]{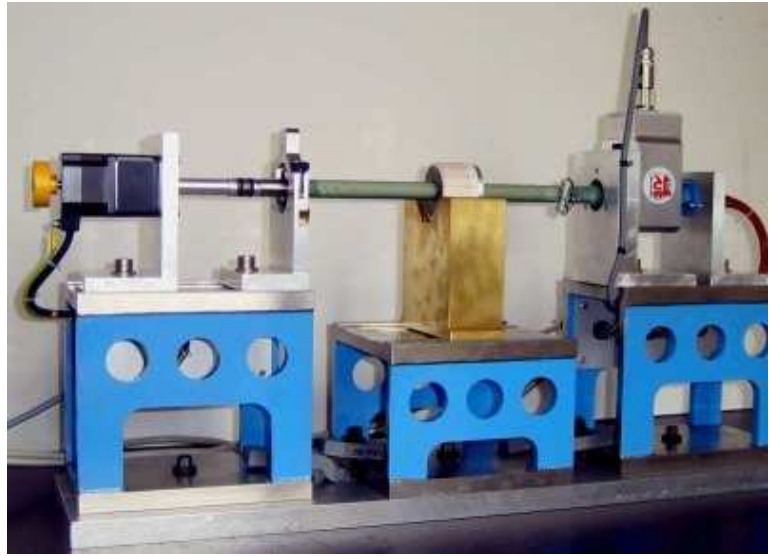} 
\caption[]{Bench measuring a small permanent quadrupole magnet for the
           drift tube of a linac. The motor is on the left and the
           encoder on the right.}
\label{fig:Linacbench} 
\end{figure}

The accuracy of finding the field direction in the coil reference
frame is high: typically better than 0.1~mrad. The difficulty is to
find the coil reference direction with respect to an external
reference. The coil average winding position is commonly misaligned by
a few mrad with respect to the coil frame. In addition, extreme care
is needed to have a perfect alignment with respect to the encoder
reference angle. An easy calibration is possible by turning end to end
either the magnet to be measured or the encoder plus coil set. It was
possible, on account of the size of the magnets to be measured, to
design the bench shown in \Fref{fig:Linacbench} for this easy
calibration.

\subsubsection{Coil rotation axis different from the magnet symmetry axis}

Similarly, the measuring coil does not necessarily rotate about the
magnet axis. The reference frame, where $z_c=0$, of the magnet axis is
usually defined by having both dipole components $B_1=A_1=0$ in a
quadrupole. With $d \cdot R_\text{ref}$ being the distance from the
rotation axis to the measuring frame, the two sets of multipoles are
related by
\begin{align}
z_m    & = z_c-d \cdot R_\text{ref}~,\label{eq:eqH15}\\
C_{m,n}& = \sum_{k=n}^\infty {\frac{(k-1)!}{(n-1)!(k-n)!} C_{c,k}} \cdot d^{k-n}~.
\label{eq:eqH16}   
\end{align}

This so-called `feed-down' correction is used when the rotation axis
cannot be accurately aligned in the magnet, for instance with a magnet
aperture small compared to the length. These magnets are measured with
`moles' travelling along the aperture~\cite{Garciamole} or coil shafts
with intermediate bearings resting in the magnet~\cite{shaft15m}.  It
must be stressed, however, that displacing the reference frame
corresponds to describing the field outside the measurement circle,
thus extrapolating the measurements outside their validity range. This
feed-down correction loses validity for large values of $d$.

\Eref[b]{eq:eqH16} can be simplified, assuming that the coil rotation
axis is near the magnet axis, i.e., $d \ll 1$:

\begin{equation}
C_{m,n}=C_{c,n} + n \cdot C_{c,n+1} \cdot d~.
\label{eq:eqH17}   
\end{equation}

The position of the rotation axis of the measuring system with respect
to the the axis of a quadrupole is therefore

\begin{equation}
(d_x + i \cdot d_y)\cdot R_\text{ref} = \frac {C_{m,1}} {C_{m,2}} \cdot R_\text{ref}~.
\label{eq:eqH18}   \end{equation}

Measuring this distance with an accuracy of 0.01\Umm{} is easily
achieved for short length measuring coils. As in \Sref{sec:Field
direction} the issue is to refer the rotation axis to the magnet
fiducials. An easy calibration is done when the magnet can be turned
upside down.

\subsection{Possible configurations according to the magnet size}
\label{sec:Possible configurations}

\Fref[b]{fig:Linacbench} shows a rotating coil bench measuring
linac permanent quadrupoles that are short and small. The motor, the
bearings of the rotating coil, and the angular encoder are accurately
aligned in a straight line with reference to the base plate.

\Fref[b]{fig:SM18} shows in comparison a 15 m long, 35 tonne LHC
superconducting dipole magnet. \Bref[b]{magneticLHC} describes the
mathematical model established on the series measurement of the LHC
superconducting magnets, giving an idea of the size of the measurement
project. The measuring shaft \cite{shaft15m} in the right part is
composed of 12 segments 1.15 m long and separated by ceramic bearings
resting on the anti-cryostat, therefore rotating off-axis by up to 3
mm with respect to the symmetry axis of the magnet. By construction,
the measuring segments are not accurately aligned azimuthally with
respect to each other. Equations~(\ref{eq:eqH14})--(\ref{eq:eqH16})
have therefore to be applied to express the results of the
measurements in the reference frame of the magnet axis.

Similarly, these LHC magnets were measured by `moles'
\cite{Garciamole} in the manufacturing industries. These moles group
in one device the coil, motor, and encoder and travel along the magnet
length pulled by a cable. Measuring the field direction with the help
of an on board inclinometer requires dedicated calibration. Note that
other techniques (\Sref{sec:sswaxis}, \cite{Garciaacmole}) can measure
the magnetic axis of quadrupole magnets.

\Fref[b]{fig:threesegments}(a) shows a coil structure designed for
easy alignment of the magnet with respect to the bearings of the
measurement system~\cite{LEPref}. This coil setup allows one to
separately align the magnetic axis on both magnet ends, then to fine
tune the alignment with the full length coil. The full length coil
gives the magnet strength, or integrated field, value directly
relevant for the accelerator beam. The magnet effective length is
obtained by calculating the central field given by the difference
between the full length coil and the sum of the end coils. The
measurement from the end coils can be compared to 3D simulation of the
magnet ends. Note that a measurement with a coil shorter than the
magnet is valid only if the longitudinal component of the field is
zero on both ends of these coils \cite{Davies92}.

\begin{figure}
\centering
\includegraphics[width=15cm]{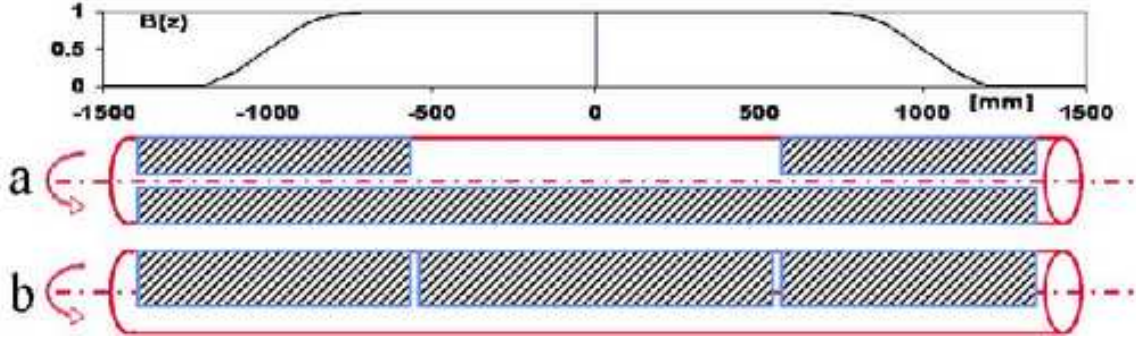} 
\caption[]{(a) Coil shaft with a full length coil and two end
           coils. (b) Separated coil structure when solution (a)
           cannot be manufactured. It implies blind spaces between the
           coils.}
\label{fig:threesegments}
\end{figure}

Small-aperture magnets render difficult the manufacture of rotating
shafts with the coils on both sides of the axis. The solution of
\Fref{fig:threesegments}(b) is preferred even though the sum of three
(or more) data sets must be done. The effective length of the blind
parts between the segments must be accurately measured and located
where the field does not vary along the axis.

% \end{document}
% HarmAccuracy.tex
\section{Accuracy limitations of the harmonic coil method} 
\label{HarmAccuracy}

Mechanical or electronic imperfections mainly degrade the measurement
of the `higher order' multipoles, \ie those with harmonic numbers
higher than the magnet multipole order. The three main error sources
will be studied in detail:
\begin{itemize}
\item voltage integrator offset coupled with irregular rotation rate of the coil,
\item error in the coil angle measurement due either to the angular
      encoder or to torsions of the coil shaft during rotation,
\item instability or movement of the rotation axis of the coil shaft
      due to gravity, bearings quality, or vibrations.
\end{itemize}

Schemes of compensation coil arrays, connected in opposition, have
been developed~\cite{Halbach1979} to remove the signal coming from the
magnet main multipole thus allowing the increase of the amplification
factor at the input of the integrator.  More importantly, these
compensation coil assemblies remove non-linear coupling coming from
the main harmonic and degrading the high-order harmonic measurement.

\subsection{Offset of the integrator coupled with varying rotation rate}
\label{sec:offset}

Stability of the rotation rate of the harmonic coil does not appear at
first order in the method. The angular fluxes, given in
Eqs.~(\ref{eq:eq6})--(\ref{eq:eqH10}) are explicitly independent of
time.  However, the voltage integrator has an offset that gives a flux
error inversely proportional to the rotation rate. The average offset
over a turn can be eliminated afterward by one of the following
methods in the case where the field is static during the coil
rotation. Obviously, a careful adjustment to zero the voltage offset
before the measurement is the most reliable method.
\begin{itemize}
\item The coil shaft can continuously rotate if the signal goes via
      slip rings from the coil to the integrator. The flux integrated
      over a full turn must be zero if the field is static. This
      formula is also valid if the current variation is constant and
      the angle zero corresponds to a zero flux in the coil, but this
      correction is sensitive to noise in the signal and could be
      detrimental. The incremental offset can be eliminated at first
      order by dividing the offset integrated over one turn by the
      number of angular intervals:
\begin{equation}
\oint{\text{Offset}} =[\Psi(2 \pi) - \Psi(0)_{\text{measured}}]~.   
\label{eq:eqA1} \end{equation}

\item Without the use of slip rings, the rotating coil has to come
      back in reverse rotation due to the instrumentation
      cable. Averaging the incremental fluxes measured during the
      forward and backward turns removes at first order the
      offset. Note that it implies as well either a constant or
      constantly changing current.
\end{itemize}

Variation of the rotation rate can further be eliminated if the time
duration of the individual angular steps are accumulated together with
the incremental fluxes and the corresponding correction is applied
before the Fourier analysis takes place. Experience has shown that the
resulting signal-to-noise enhancement is limited by imperfect removal
of this effect, even for rotation rates measured constant within a few
per cent.

We will see in \Srefs{sec:dipcomp} and \ref{sec:quadcomp} that most of
these degradations of the high order multipole measurement due to
electronics can be reduced by the use of compensation coil schemes.

\subsection{Measurement errors related to the angle and encoder}
\label{sec:encoder}

The accuracy in measuring the harmonic coefficients of the field
depends directly on the angular precision of the integrator
triggers. The encoder quality, the torsion of the frame holding the
coil or linking it to the encoder must all be
considered. \Bref[b]{Davies92} gives the formalism to calculate these
effects taking into account all kinds of torsions and vibrations. The
case of a periodic error when measuring a pure dipole magnet will be
detailed here. General equations can be found in \Bref{JainCAS98}.

An angular misalignment between the axis of the encoder and the axis
of the coil can lead to a difference between the measured and real
angular position of the coil. This error is best approximated by a
sine function. A torque on the bearing of the encoder due to a
parallel misalignment between these axes gives a similar error of
amplitude defined as $\epsilon$:
\begin{equation}
\theta_\text{meas}=\theta + \epsilon \cdot \sin\theta~.   
\label{eq:eqA2} 
\end{equation}

A perfect dipole aligned with respect to the measuring equipment is
defined by $B_1 = 1$, all other coefficients $B_n , A_n = 0$. As the
angular encoder errors are supposed to be small, a first-order
estimation gives
\begin{equation}
\Psi(\theta) \propto \cos(\theta_\text{meas})
  ~~=~~ \cos(\theta + \epsilon \cdot \sin\theta) 
  ~~=~~ \cos(\theta)- \frac{\epsilon}{2} \cdot (1-\cos 2\theta)~.   
\label{eq:eqA3} \end{equation}

An erroneous quadrupole term, not present in the magnet, will show
itself in the Fourier analysis of the integrated
voltage. \Fref[b]{fig:errorencoder} represents it. Its value is
related to the amplitude of the error of the encoder by
\begin{equation}
\delta B_2/ B_1 ~~=~~ \delta b_2
                ~~ =~~ \frac{\epsilon}{2} \cdot \frac{K_1}{K_2}~.   
\label{eq:eqA4} 
\end{equation}

\begin{figure}
\begin{center}
\includegraphics[width=125mm]{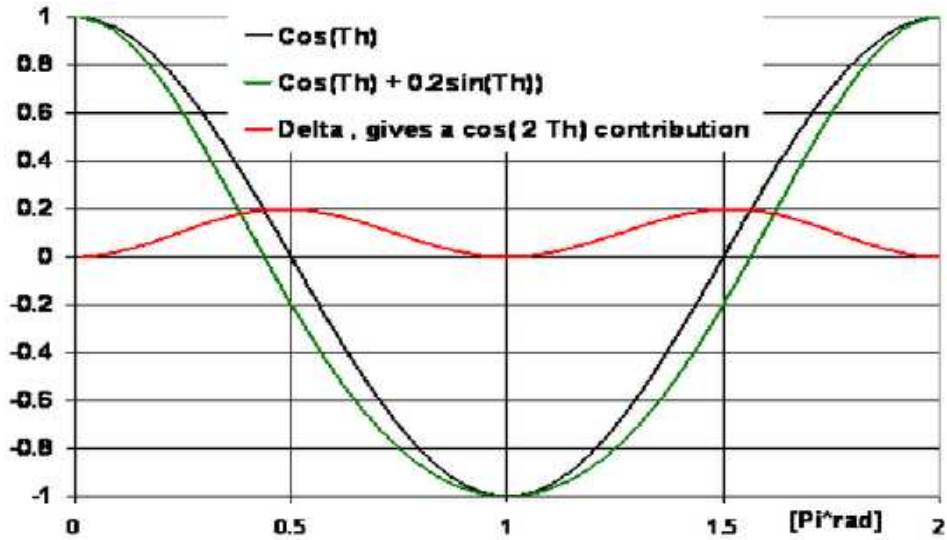}
\end{center}
\caption[]{Flux measured in a dipole with a perfect angle measurement
           and a first-order error. Their difference gives an
           erroneous quadrupole term.}
\label{fig:errorencoder}
\end{figure}
In order to give orders of magnitude, \Eref{eq:eqA4} can be further
simplified with the following hypothesis for the radii sketched in
\Fref{fig:simplecoil}:
\begin{equation*}
R_1  = 0~, \quad
R_2  = R_\text{ref}~.
\end{equation*}

The erroneous quadrupole $\delta b_2 = 10^{-3}$, \ie 10 units for an
angular error $\epsilon = 1\Umrad$ since the ratio of the sensitivity
factors is
\begin{equation}
\frac{K_n}{K_1} = \frac{1}{n}\cdot \left(\frac{R_2}{R_\text{ref}}\right)^{n-1} 
                = \frac{1}{n}~.     
\label{eq:eqA42} 
\end{equation}

More generally, any systematic angular error can be decomposed in a Fourier series:
\begin{equation}
\theta_\text{meas}=\theta + \sum_k \gamma_k\cos k\theta + \epsilon_k\sin k\theta~.     
\label{eq:eqA5} 
\end{equation}

These $\gamma_k$ and $\epsilon_k$ will generate erroneous normal and
skew multipoles when measuring a pure dipole magnet:
\begin{align}
\delta b_n  & = \frac{nK_n}{2K_1} (\epsilon_{n-1} +  \epsilon_{n+1})~, \nonumber\\
\delta a_n  & = \frac{nK_n}{2K_1} (\gamma_{n-1} +  \gamma_{n+1})~.
\label{eq:eqA7} 
\end{align}

\subsection{Removal of the main harmonic in a dipole magnet
            by using compensation coil scheme}
\label{sec:dipcomp}

The previous section shows that high precision in the measurement of
the multipoles requires high quality encoders. In addition, the
geometry of the equipment to be built, namely the ratio between magnet
aperture and length, can lead to severe difficulties in achieving a
torsion-free frame holding the coils.

\begin{figure}
\centering
\includegraphics[width=125mm]{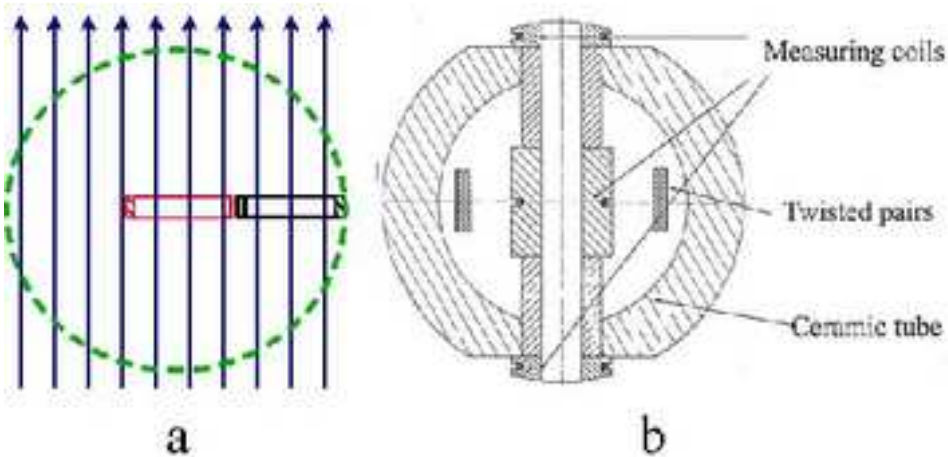} 
\caption[]{(a) Assembly of two radial coils with same effective
           surface connected in opposition to remove the main dipole
           component from the signal. (b) Cross-section of the three
           tangential coils as used in \Bref{magneticLHC}. One of the
           external coils is kept as a spare.}
\label{fig:dipcomp}
\end{figure}

A suitable geometry for the measuring-coil system may be chosen that
reduces the angular precision needed. Two coils having the same
sensitivity with respect to the dipole term (the same surface) are
electrically connected in opposition at the input of the
integrator. The resulting sensitivity factor as calculated in
\Eref{eq:eqH11} becomes $K_1 = 0$. Their radial positions are sketched
in \Fref{fig:dipcomp}. Consequently, the dipole harmonic is rejected
in the flux measured. Rejection ratios of 300 to 2000 can be achieved
by sorting the coils according to their effective area and by
adjusting them to be parallel. This assembly of coils is sometimes
called `bucking coils' in the literature.

\subsection{Measurement errors related to coil transverse movement}
\label{sec:transversemove}

In a dipole magnet, the measuring coil assemblies of
\Fref{fig:dipcomp}, displaced laterally or vertically, will induce no
voltage on the integrator connected to it. This is not true if higher
harmonics are present or when a quadrupole or sextupole magnet has to
be measured. A spurious signal therefore shows up if the mechanics is
such that the harmonic coil does not describe a perfect circle when
rotating. Calculations show that without proper rejection of the main
harmonic, one cannot measure with high accuracy the field quality of a
quadrupole.

\begin{figure}
\centering
\includegraphics[width=12cm]{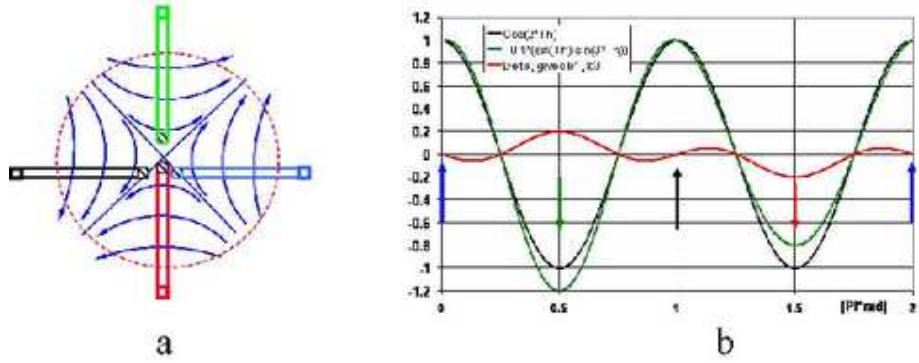} 
\caption[]{(a) A non-rigid coil shaft moving up and down, due to
           gravity, during rotation in a quadrupole field. (b) The
           flux measured with a perfect rotation and a first-order
           error. Their difference gives erroneous dipole and
           sextupole terms.}
\label{fig:transversemove}
\end{figure}

The deflection of the frame holding the coil gives a simple measure of
this effect. The hypothesis is made that the coil is wound on a flat
plate that bends due to its own weight when horizontal, and is
straight when vertical [\Fref{fig:transversemove}(a)]. The centre
of the coil consequently moves up and down twice per turn. This
vertical displacement of amplitude $d \cdot R_\text{ref}$ is
approximated by
\begin{equation}
\text{Displ.} = i \cdot d \cdot R_\text{ref} \cdot \cos(2\theta)~.
\label{eq:eqA8} 
\end{equation}

The bounds of the integral of \Eref{eq:eqH8} become with the
hypothesis of a simple coil [given by \Eref{eq:eqA42}]
\begin{align}
R_2(\theta) &= R_\text{ref} \cdot (i \cdot d \cdot \cos(2\theta) 
                                   + e^{i\theta})~,\nonumber \\
R_1(\theta) &= R_\text{ref} \cdot i \cdot d \cdot \cos(2\theta)~.
\label{eq:eqA9} 
\end{align}

These lateral movements during coil rotation induce non-linear
coupling with the quadrupolar field and generate erroneous multipoles
of lower and higher orders. \Fref[b]{fig:transversemove}(b) sketches
the curves giving the angular flux with and without this vertical
movement. A first-order calculation, i.e., neglecting terms in $d^2$,
into a pure quadrupolar field (only $B_2 \neq 0$) gives the following
erroneous skew components:
\begin{align}
\delta B_1/B_2 = \delta b_1 & =d \\ 
                 \delta b_3 & =-3d~.
\label{eq:eqA10} 
\end{align}

A deflection due to gravity of $0.02\Umm$ with $R_\text{ref} = 20\Umm$
(\ie $d = 0.001$) generates an erroneous dipole corresponding to 0.02
mm of axis displacement and a relative sextupole $b_3 = 0.003$ (or 30
units). This simple calculation therefore stresses the importance of
designing a measuring shaft to hold the rotating coils as rigidly as
possible, and having as smooth as possible rotation with the help of
high-quality bearings and motorisation.

The formal development of the erroneous components created by any
transverse movement in any real magnet can be found in
\Brefs{Davies92}--\!\cite{JainCAS98}.

\subsection{Removal of the main harmonic in a quadrupole magnet 
            by using compensation coil scheme} 
\label{sec:quadcomp}

A coil scheme for an effective removal of the main quadrupole harmonic
has to be insensitive to both angular and transverse movements in a
pure quadrupole, i.e., to both the quadrupole and dipole sensitivity
factors as defined in \Eref{eq:eqH11}:
\begin{equation}
K_1=K_2=0~.
\label{eq:eqA11} 
\end{equation}

\begin{figure}
\centering
\includegraphics[width=\linewidth]{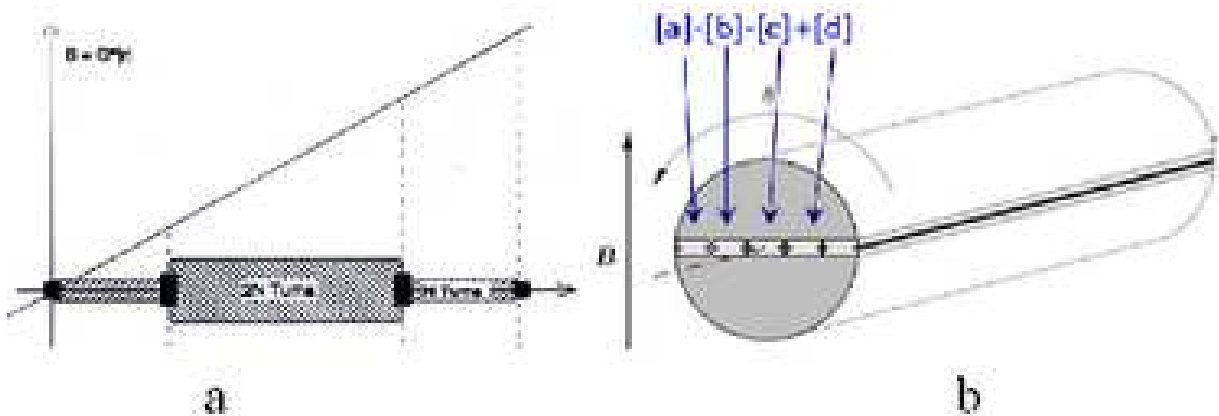} 
\caption[]{Two geometries to compensate the dipole and quadrupole
           components: (a) two coils having the same effective
           surface, (b)~four equal radial coils give the same
           compensation, one of the external coils is kept as a
           spare.}
\label{fig:quadcomp}
\end{figure}

The two schemes of \Fref{fig:quadcomp} fulfil these two conditions. 

It is difficult to have exactly the same effective area for the two
coils of the scheme of \Fref{fig:quadcomp}(a), i.e., the inside coil
having half the width and twice the number of turns of the outside
one. This scheme induces in addition a high loss of sensitivity to
measure the low order multipoles, in particular the
sextupole. Although such a coil assembly rejects correctly the
quadrupole harmonic, a factor of more than five is lost in sensitivity
to measure the sextupole term compared to the sensitivity given by the
single external coil. Since this coil assembly is insensitive to any
lateral displacement in a quadrupole, turning about another axis will
maintain the rejection. A maximum value for the sextupole sensitivity
factor $K_3$ is obtained with the radii of the external coil being
\cite{WalckiersCAS92}:
\begin{align}
R_2 &= R_\text{ref} ,  \nonumber \\
R_1 &= -1/2 \cdot R_\text{ref}~.
\end{align} 

\begin{figure}   
\centering
\includegraphics[width=15cm]{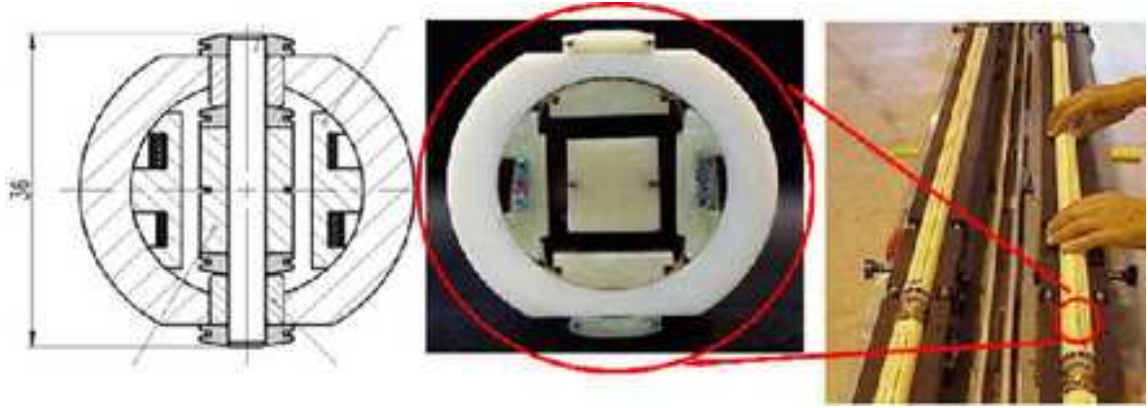} 
\caption[]{A compensation coil scheme to measure quadrupoles. Four
           tangential coils are used to remove the dipole and
           quadrupole components, one of the external coils is kept as
           a spare.}
\label{fig:QuadrCeramicCoil}   
\end{figure}

The other scheme is based on four equal coils giving the same
geometrical properties [\Fref{fig:quadcomp}(b) and
\Fref{fig:QuadrCeramicCoil}]. A rejection ratio of more than 100 can
be achieved by sorting and careful positioning of the four coil
assemblies. Note that these two types of coil assemblies can be used
to measure dipole magnets according to \Sref{sec:dipcomp} by using
only the central and external coils.

\section{Calculation of the sensitivity factors of the harmonic coils}
\label{sec:coil}
 
\subsection{Tangential coil sensitivity}
\label{sec:tangentialcoils}

Tangential measuring coils are preferred for easier manufacture and
better stability of the rotating coil frame
(\Sref{sec:quadcomp}).

In order to calculate the sensitivity of tangential coils,
\Erefs{eq:eqH8}--(\ref{eq:eqH11}) have to be generalized by
putting $R_1$ and $R_2$ as 2D variables in the complex plane:
\begin{equation}
K_n=N_t\cdot L \cdot \frac{Z_2^n-Z_1^n}{n\cdot R_r^{n-1}}~.
\label{eq:eqK1}   \end{equation}

\begin{figure}
\centering
\includegraphics[width=12cm]{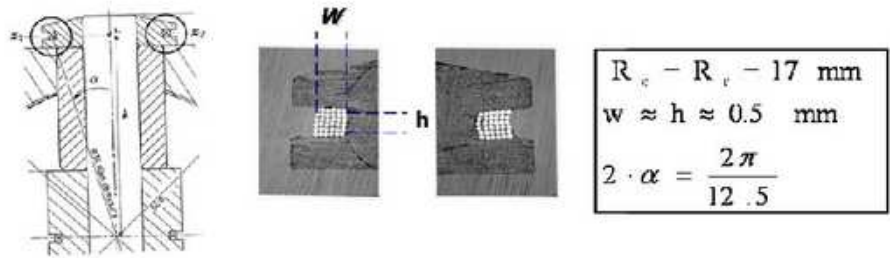} 
\caption[]{Parameters of the tangential coil cross-section used for
           the series measurements of the LHC magnets}
\label{fig:tancoil}
\end{figure}
The tangential coil of \Fref{fig:tancoil} is horizontal on the $y_+$
for the first point of rotation and gives real expressions for the
$K_{2n+1}$. The normal components of the multipoles belonging to the
dipole symmetry have indeed a vertical field on the $y$ axis. The
normal components of the quadrupole, octupole, \etc have an
horizontal field on the $y$ axis and the $K_{2n}$ factors have
imaginary values. \Eref[b]{eq:eqK1} is simplified to
\begin{equation}
K_n=-2\cdot i^{n+1}\cdot N_t\cdot L \cdot \frac{R_c^n\sin(n\alpha)}{n\cdot R_r^{n-1}}~,
\label{eq:eqK3}   
\end{equation}
with the two winding positions
\begin{eqnarray}
Z_1=R_c\cdot e^{i(\pi /2 + \alpha)}~,\nonumber\\
Z_2=R_c\cdot e^{i(\pi /2 - \alpha)} \nonumber.
\end{eqnarray}

\Eref[b]{eq:eqK3} shows that tangential coils are insensitive to
multipole order having an angular period, or an integer number of
periods, corresponding to the opening angle.  This drawback is easily
overcome by having the opening angle either small enough or
corresponding to
\begin{equation}
2\alpha =\frac{2\pi}{n+1/2}~.
\label{eq:eqK3a}   
\end{equation}

Note that the $K_n$ would all be imaginary if the tangential coil
starts to rotate in vertical position centred on the $x_+$ axis since
the field of all normal components (the $B_n$) is vertical on the $x$
axis. \Eref[b]{eq:eqK1} gives with this hypothesis
\begin{equation}
K_n=-2\cdot i\cdot N_t\cdot L \cdot \frac{R_c^n \sin(n\alpha)}{n\cdot R_r^{n-1}}~.
\label{eq:eqK4}   
\end{equation}

\Eref[b]{eq:eqK1} is valid for any coil and should be used in the case
of compensation coil schemes (\Srefs{sec:dipcomp}--\ref{sec:quadcomp})
where the relative angle between the different coils should be taken
into account. General expressions for any ill-positioned coil can be
found in \Bref{Deniau}.

\subsection{The coil winding is not point-like}
\label{sec:calccoilfactor}

\Erefs[b]{eq:eqH11} and (\ref{eq:eqK1}) are valid for coils having
point-like windings.  The windings are usually multi-turn and their
dimensions, width and height in \Fref{fig:tancoil}, should be taken
into account by summing the contribution of the position of each
individual turn. This summation can be simplified by a normal integral
over the section of the winding with the assumption that the density
of the turns is homogeneous over this section.  The point-like radius
of \Eref{eq:eqK1} has to be replaced by the $n$-order average over the
winding section $S$~\cite{Deniau}:
\begin{equation}
\left\langle Z^n \right\rangle = \frac{\int_S {z^n \cdot \rmd z}}{S} ~.
\label{eq:eqK5}   
\end{equation}

\begin{figure}
\centering
\includegraphics[width=13cm]{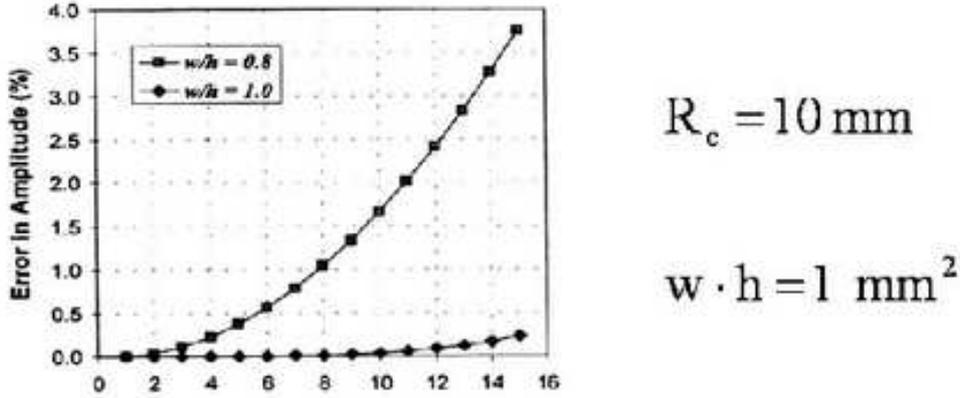} 
\caption[]{Error on the sensitivity factor $K_n$ as a function of the
           multipole order if the winding is considered point-like for
           a square and rectangular cross-section (courtesy of A. Jain)}
\label{fig:squaresection}
\end{figure}

The importance of the corresponding correction is, however,
small. Their difference with the point-like case grows with the
multipole order whilst the multipole perturbations they measure
generally decrease with the order. In addition, the corrections to
apply become negligible for square-shaped windings even for high
harmonic numbers. This is demonstrated by \Fref{fig:squaresection}
extracted from \Bref{JainCAS98}. A coil winding of $1\Umm^2$ located
at $10\Umm$ radius and calculated with the point-like approximation
would give an error of two per mille for the 30-pole ($n=15$). These
effects should therefore be estimated only for measuring coils with
aspect ratios far from a square shape like coils based on printed
circuits.

% HarmExperience.tex
\section{Experience with the harmonic coil method}
\label {sec:HarmExperience}

\subsection{Precision of the measurement of the main field component}

The measurement of the field strength integrated over the magnet
length, or of the transfer function defined as this integral divided
by the current, is usually insensitive to electronic or mechanical
noise sources like the shaft rigidity, smoothness of the rotation,
vibrations. However, it depends strongly on the calibration of the
coil surface for the dipole measurement and the positioning of the
coils in the coil frame for quadrupoles and higher order
magnets. Techniques for calibrating the coils are described elsewhere
in these proceedings. They can be based on measurements with the
single stretched wire technique described in
\Sref{sec:sswdipole}. Accurate measurements of the gains of the
electronics and of the excitation current are obviously key
elements. Relative precisions of $10^{-4}$ are obtained to measure
dipole magnet strengths and 3 to 10 times worse for quadrupoles with
careful calibrations of all elements \cite{smirnov2006}.

\subsection{Measuring the higher-order multipoles}

The current stability must be extremely high over the turn duration,
in the range of ppm to tens of ppm, in order not to create spurious
high-order multipoles if measured with a single coil. It is possible
to acquire the current value for each incremental flux and apply a
first-order correction but this requires a high bandwidth and
high-quality measurement.

Experience indicates that the low-order multipoles are more sensitive
to the equipment imperfections analysed: voltage integrator offset,
rigidity of the frame, irregularities in the rotation giving both
torsional and lateral movements. In other words, the sextupole is the
difficult multipole to measure in a quadrupole, and that justifies a
careful design of the compensation coil scheme to increase $K_3$ as
explained in \Sref{sec:quadcomp}. Repeatability of 0.01 unit (or
1~ppm relative to the main component) has been achieved in the best
cases \cite{harmaccuracy}.

The use of compensation schemes improves the measurement of
higher-order multipoles for the following reasons.
\begin{itemize}
\item The measured signal has a smaller amplitude and can therefore be
      amplified, reducing electronic noise. It also reduces the errors
      coming from the offset and the non-linear coupling between
      offset, rotation rate variations, and signal of the main order
      harmonic.
\item Excitation current variations, including the ripple from the
      magnet power supply, also couple with the main order signal
      which is removed with compensation coils. For the same reason,
      measurements with slowly ramping currents are improved.
\item An improvement proportional to the rejection ratio is obtained
      for all noises generated by mechanical instabilities:
      displacement of the rotation axis of the coil shaft due to
      gravity, quality of the bearings, or vibrations.
\end{itemize}

\subsection{Measurement of pulsed magnets with harmonic coils}

Measurement of pulsed magnets, i.e., excitation current changing with
time, is hardly covered by the rotating coil method. The measurement
of the magnet strength, i.e., the main multipole, is spoiled by the
field change over the duration of the coil revolution. Accurate
synchronization for the measurement of the excitation current at each
angular position could be mandatory to get enough precision. Obviously,
increasing the rotation rate of the coil shaft and the bandwidth of
the electronics greatly helps and must be taken into account at the
design phase of the rotating coils, \ie the highest number of turns
for the coil windings is not necessarily the optimum. As an example,
the $15\Um$ long coil shaft of \Fref{fig:SM18} currently measures, at
8 turns per second, the field quality of the LHC magnets with high
temporal accuracy.

The measurement of the field quality, i.e., the higher-order
multipoles, is greatly improved \cite{multipoleramp} in this case by
the help of a coil scheme to compensate the main multipole
(\Srefs{sec:dipcomp} and \ref{sec:quadcomp}).

For fast ramping magnets, it is possible to measure flux variation
between zero and nominal current with the harmonic coil fixed at a
sufficiently large number of angular positions to perform the Fourier
analysis. Experimental work is going on at GSI \cite{Schnizer2008} for
the FAIR project and at CERN for the new Linac quadrupole magnets
having a duty cycle of a few milliseconds. This method requires an
accurate angle positioning, obtained for instance by a high-resolution
stepping motor. It would bring the advantages of the Morgan coil
\cite{Morgan} technique with a simplified coil frame to be
manufactured.

% \end{document}
% Compare.tex
\section{Conclusion: compare the different measurement techniques}

Coils and stretched wires are tools particularly adapted to
measurement of accelerator magnets. They give reduced values for
parameters controlling particle beams: cross-section components of the
field integrated along the magnet length.

The SSW based system can be considered reference equipment to measure
in a static field dipole and quadrupole strength, \ie main field
components, where precisions of $10^{-3}$ to $10^{-4}$ are
needed. These instruments measure with high accuracy the main field
direction and magnetic axis of quadrupole and sextupole magnets. They
are being used more and more for these measurement goals in particular
with the requirement to better align magnets for synchrotron light
sources and with the need for smaller and smaller apertures for
high-energy physics accelerators.

Flip coils are a reduced technique of the harmonic rotating coil
method. The theory associated with their use is, however, mandatory to
correctly measure fast pulsed and curved magnets.

The rotating coil method is a general and accurate method to measure
the field quality of magnets: integrated field value, higher order
multipoles, and magnetic axis. Recent instrumentation and acquisition
systems allow high bandwidth and fully automated measurements. This
method is the obvious choice for normal quadrupole magnets, and for
superconducting magnets having circular apertures and where beam
optics considerations require unprecedented precision in the field
quality.

These various methods complement each other. They are complemented by
the use of Hall plates for local measurements and of NMR based
instruments for high absolute accuracy and calibration. A cross-check
between these various methods should be used whenever possible to
ascertain precision in magnet measurements.
\section*{Acknowledgements}

The sections related to the single stretched wire are mostly extracted
from the experimental and theoretical work done in different
laboratories, in particular by J. Dimarco, A. Jain, A. Temnykh, and
Z. Wolf. The list of names of my numerous CERN colleagues who are at
the basis of my competence in the field of magnetic measurements is
too long to mention here.
% Bibliography.tex

% Bibliography.tex

\end{document}